\documentclass[trackchanges]{aastex701}

\usepackage[]{subfigure}
\usepackage{amsmath}
\usepackage{booktabs}
\usepackage{comment}

\begin{document}
\title{IXPE view of the Crab pulsar following the 17 July and 6 August 2025 glitches}

\author[0000-0002-7781-4104]{Paolo Soffitta}
\affiliation{INAF Istituto di Astrofisica e Planetologia Spaziali, Via del Fosso del Cavaliere 100, 00133 Roma, Italy}
\email{paolo.soffitta@inaf.it}

\author[0000-0002-8848-1392]{Niccol\`{o} Bucciantini}
\affiliation{INAF Osservatorio Astrofisico di Arcetri, Largo Enrico Fermi 5, 50125 Firenze, Italy}
\affiliation{Dipartimento di Fisica e Astronomia, Universit\`{a} degli Studi di Firenze, Via Sansone 1, 50019 Sesto Fiorentino (FI), Italy}
\affiliation{Istituto Nazionale di Fisica Nucleare, Sezione di Firenze, Via Sansone 1, 50019 Sesto Fiorentino (FI), Italy}
\email{niccolo.bucciantini@inaf.it}

\author[0000-0001-6395-2066]{Josephine Wong}
\affiliation{Department of Physics and Kavli Institute for Particle Astrophysics and Cosmology, Stanford University, Stanford, California 94305, USA}
\email{joswong@stanford.edu}

\author[0000-0001-5848-0180]{Denis Gonz\'{a}lez-Caniulef}
\affiliation{IRAP, CNRS, 9 avenue du Colonel Roche, BP 44346, F-31028 Toulouse Cedex 4, France}
\email{denis.gonzalez-caniulef@irap.omp.eu}

\author[0000-0002-4576-9337]{Matteo Bachetti}
\affiliation{INAF Osservatorio Astronomico di Cagliari, Via della Scienza 5, 09047 Selargius (CA), Italy}
\email{matteo.bachetti@inaf.it}

\author[0000-0003-1074-8605]{Riccardo Ferrazzoli} 
\affiliation{INAF Istituto di Astrofisica e Planetologia Spaziali, Via del Fosso del Cavaliere 100, 00133 Roma, Italy}
\email{riccardo.ferrazzoli@inaf.it}

\author[0000-0002-0105-5826]{Fei Xie}
\affiliation{Guangxi Key Laboratory for Relativistic Astrophysics, School of Physical Science and Technology, Guangxi University, Nanning 530004, China}
\affiliation{INAF Istituto di Astrofisica e Planetologia Spaziali, Via del Fosso del Cavaliere 100, 00133 Roma, Italy}
\email{xief@gxu.edu.cn}
\author[0000-0003-4925-8523]{Enrico Costa}
\affiliation{INAF Istituto di Astrofisica e Planetologia Spaziali, Via del Fosso del Cavaliere 100, 00133 Roma, Italy}
\email{enrico.costa@inaf.it}
\author[0000-0001-7397-8091]{Maura Pilia}
\affiliation{INAF Osservatorio Astronomico di Cagliari, Via della Scienza 5, 09047 Selargius (CA), Italy}
\email{maura.pilia@inaf.it}

\author[0000-0003-3842-4493]{Nicol\`{o} Cibrario}
\affiliation{Istituto Nazionale di Fisica Nucleare, Sezione di Torino, Via Pietro Giuria 1, 10125 Torino, Italy}
\affiliation{Dipartimento di Fisica, Università degli Studi di Torino, Via Pietro Giuria 1, 10125 Torino, Italy}
\email{niccolo.cibrario@to.infn.it}

\author[0000-0002-6401-778X]{Jack T. Dinsmore}
\affiliation{Department of Physics and Kavli Institute for Particle Astrophysics and Cosmology, Stanford University, Stanford, California 94305, USA}
\email{jtd@stanford.edu}

\author[0000-0002-7574-1298]{Niccol\`{o} Di Lalla}
\email{niccolo.dilalla@stanford.edu}
\affiliation{W. W. Hansen Experimental Physics Laboratory (HEPL)}
\affiliation{Kavli Institute for Particle Astrophysics and Cosmology (KIPAC)}
\affiliation{Department of Physics and SLAC National Accelerator Laboratory, Stanford University, Stanford, CA 94305}

\author[0000-0002-3638-0637]{Philip Kaaret}
\affiliation{NASA Marshall Space Flight Center, Huntsville, AL 35812, USA}
\email{philip.kaaret@nasa.gov} 

\author[0009-0007-8686-9012]{Kuan Liu}
\affiliation{Guangxi Key Laboratory for Relativistic Astrophysics, School of Physical Science and Technology, Guangxi University, Nanning 530004, China}
\email{liuk@st.gxu.edu.cn}

\author[0000-0001-7263-0296]{Tsunefumi Mizuno}
\affiliation{Hiroshima Astrophysical Science Center, Hiroshima University, 1-3-1 Kagamiyama, Higashi-Hiroshima, Hiroshima 739-8526, Japan}
\email{mizuno@astro.hiroshima-u.ac.jp}

\author[0000-0002-6548-5622]{Michela Negro}
\affiliation{Department of Physics and Astronomy, Louisiana State University, Baton Rouge, LA 70803 USA}
\email{michelanegro@lsu.edu}

\author[0000-0002-5847-2612]{C.-Y. Ng}
\affiliation{Department of Physics, The University of Hong Kong, Pokfulam, Hong Kong}
\email{ncy@astro.physics.hku.hk}

\author[0000-0002-5448-7577]{Nicola Omodei}
\affiliation{W. W. Hansen Experimental Physics Laboratory (HEPL)}
\affiliation{Kavli Institute for Particle Astrophysics and Cosmology (KIPAC)}
\affiliation{Department of Physics and SLAC National Accelerator Laboratory, Stanford University, Stanford, CA 94305}
\email{nicola.omodei@stanford.edu}

\author[0009-0004-5622-1854]{Simone Pagliarella}
\affiliation{INAF Istituto di Astrofisica e Planetologia Spaziali, Via del Fosso del Cavaliere 100, 00133 Roma, Italy}
\affiliation{Dipartimento di Fisica, Università degli Studi di Roma “La Sapienza”, Piazzale Aldo Moro 5, I-00185 Roma, Italy}
\affiliation{Dipartimento di Fisica, Università degli Studi di Roma “Tor Vergata”, Via della Ricerca Scientifica 1, I-00133 Roma, Italy}
\email{simone.pagliarella@inaf.it}

\author[0000-0002-8665-0105]{Stefano Silvestri}
\affiliation{Dipartimento di Fisica, Universit\`{a} di Pisa, Largo B. Pontecorvo 3, 56127 Pisa, Italy}
\affiliation{Istituto Nazionale di Fisica Nucleare, Sezione di Pisa, Largo B. Pontecorvo 3, 56127 Pisa, Italy}
\email{stefano.silvestri@pi.infn.it}

\author[0000-0001-9108-573X]{Yi-Jung Yang}
\affiliation{Center for Astrophysics and Space Science (CASS), New York University Abu Dhabi, PO Box 129188, Abu Dhabi, UAE}
\email{yy6063@nyu.edu}

\author[0000-0002-5270-4240]{Martin C. Weisskopf}
\affiliation{NASA Marshall Space Flight Center, Huntsville, AL 35812, USA}
\email{Martin.C.Weisskopf@nasa.gov}

\begin{abstract}

The Crab pulsar experienced two relatively small glitches separated by only 20 days in September and October 2025. IXPE observed the source twice, with delay times since the glitch epoch ranging between 35 and 75 days, depending on the observation. We carried out a multi-method analysis to investigate whether there is evidence for significant changes in the polarization properties of the pulsar, underlying possible variations in the pulsar magnetosphere itself following the glitches. Specifically, we performed: (1) phase-averaged polarimetry of the Crab pulsar before and after the glitches, following an approach similar to that adopted in 2019 by PolarLight, a non-imaging CubeSat-class photoelectric polarimeter which observed a change in the X-ray polarization within 100 days after a stronger glitch in July 2019; (2) a comparison, before and after the glitch, of phase-resolved X-ray polarimetry with IXPE, not possible with PolarLight. Furthermore, we investigated, by means of phase-resolved optical (OPTIMA) polarimetry, whether a significant change in the X-to-optical lag was present in the data before and after the glitch. We find no evidence of a change in the polarization for the pulsar emission before and after the glitch, We use the upper limits obtained to estimate the maximum change in magnetic obliquity allowed by the data, using the standard rotating vector model and assuming that the glitch is due to a neutron-star quake. We constrain this maximum change to be no greater than $\pm4^{o}$ at the 95\% confidence level.

\end{abstract}



\section{Introduction} 
Glitches are abrupt spin-up events, sudden increases in spin frequency $\Delta\nu>0$, that interrupt the secular spin-down of pulsating neutron stars powered by rotation (radio/optical/X-ray pulsars) or by strong magnetic fields (magnetars), \citep[see e.g.][]{Fuentes2017,Kaspi2017}. 
Although the first pulsar glitch was discovered in the Vela pulsar nearly 60 years ago \citep{Radhakrishnan1969b}, the physical mechanism responsible for these events remains under investigation and is not yet firmly established. Elastic stress in the solid crust and a subsequent starquake may reduce the moment of inertia, spinning up the star \citep{Baym1969}.  Another model relies on the different physical status of the neutron star solid crust at the surface and superfluid at the interior with vortexes pinned on the solid crust. When vortexes un-pin, they transfer angular momentum to the solid crust provoking a glitch \citep{Anderson1975}. Post-glitch recovery is modeled in the vortex–creep framework \citep{Alpar1984} as the thermally activated resumption of vortex creep in pinned superfluid regions after the glitch.  
Although starquake-driven scenarios allow for a glitch to shear the  foot-points of magnetic field lines, and  to give rise to a magnetospheric twist that later relaxes toward a predominantly dipolar configuration \citep{Beloborodov2009,Antonopoulou+15a,Akbal+15b}, within the vortex-creep framework the glitch is primarily an internal exchange of angular momentum between a pinned superfluid and the crust and, therefore, does not require any change in the external magnetosphere (\citet{Gugercinolu17} and for a review see \cite{Haskell2015}
and \cite{Ruderman1998}).\\
\\
A novel way to probe the magnetic field configuration of pulsars and to gain insight into the emission mechanisms and sites is now available thanks to the advances in X-ray polarimetry made possible by the Imaging X-ray
Polarimetry Explorer (IXPE). Launched on 9 December 2021, it carries a polarization-sensitive
imaging instrument based on the Gas Pixel Detector (GPD) specifically developed for this purpose \citep{Weisskopf2022,Soffitta2021,Baldini2021}.    
Before IXPE’s launch, a Gas Pixel Detector (GPD) was flown as the payload (named Polar-Light) of a CubeSat \citep{Feng2019}. This single GPD, operated with a micropore (MPO) collimator (hence non-imaging) to minimize X-ray background. It observed several bright sources
with a small effective area (less than $1~\mathrm{cm}^2$) over weeks–months integrations, including the Crab \citep{Feng2020b,Long2021}, Sco~X-1 \citep{Long2022}, and A0535+26 \citep{Long2023}. In particular, for the Crab, PolarLight provided modern spaceborne X-ray polarimetry and reported polarization properties, and their temporal evolution across the 2019 glitch \citep{Shaw2019}, consistent with and extending historical measurements \citep{Feng2020b,Long2021}.
The X-ray polarization fraction (PF), measured in the
\emph{pulse-on} phase (Main Pulse (MP) + Interpulse (IP)), decreased from the pre-glitch value
$\mathrm{PF}_{\mathrm{on,pre}} = 0.288^{+0.071}_{-0.073}$ with polarization angle
$\mathrm{PA}_{\mathrm{on,pre}} = 142.7^{\circ} \pm 7.2^{\circ}$ to the post-glitch value
$\mathrm{PF}_{\mathrm{on,post}} = 0.101^{+0.047}_{-0.051}$ with
$\mathrm{PA}_{\mathrm{on,post}} = 153.0^{\circ} \pm 14.4^{\circ}$.
The ``post'' measurements span the 100 days following the glitch, whereas the pre-glitch measurements were taken shortly before the event and approximately 100 days earlier.
In the \textit{pulse-off} phase (defined as in OSO-8) the polarization properties were found constant in time  since before  the glitch  with $\mathrm{PF}_{\mathrm{off,pre}} = 0.137^{+0.076}_{-0.110}$ and polarization angle  $\mathrm{PA}_{\mathrm{off,pre}} = 149.9^{\circ} \pm 21.0^{\circ}$ to the post-glitch values $\mathrm{PF}_{\mathrm{off,post}} = 0.127^{+0.061}_{-0.067}$ and polarization angle  $\mathrm{PA_{\mathrm{off,post}} = 138.7^{\circ} \pm 15.0^{\circ}}$.
The pulse-on change is significant at the $\sim 3\sigma$ level across multiple analysis
methods, including Bayes factors, Bayesian posterior inference, and bootstrap resampling \citep[see ][]{Long2021}. This is consistent with (and may suggest) a modification of the pulsar’s magnetosphere following the glitch.  Notably, the \emph{nebula-only} polarization angle (PA) measured by PolarLight and by IXPE
\citep{Bucciantini2023} is lower than the historical OSO-8 value by $\sim 18$–$21^{\circ}$ (i.e., $\Delta\mathrm{PA}\simeq -18^{\circ}$ to $-21^{\circ}$). A plausible interpretation is that the large-scale field geometry in the inner nebula has evolved over time. 

Unfortunately, separating the pulsar polarization signal, with a high statistical confidence, from the underlying nebula, in space-integrated polarization measures is quite demanding, and temporal variability of the nebular properties, can easily introduce substantial uncertainties. Thanks to its X-ray optics \citep{Ramsey2022} and the spatial resolution of the detector \citep{Soffitta2013,Fabiani2014,Ferrazzoli2025}, IXPE can mitigate nebula contamination in pulsar measurements. Unlike PolarLight, IXPE, because of its much larger effective area and imaging, can moreover perform phase-resolved polarimetry of the Crab pulsar, separating MP and IP. This is central to the comparison of IXPE and PolarLight results, given the stronger 2019 glitch. This increased performance in extracting the pulsar signal has been achieved via: (i) imaging selection combined with phase-resolved analysis (e.g., \cite{Bucciantini2023}); (ii) simultaneous fitting of the pulsar light
curve and the Pulsar Wind Nebula (PWN) surface-brightness distribution constrained by Chandra imaging \citep{Wong2023,Wong2024}.

As an example of the superior capabilities of IXPE, based on OPTIMA optical results \citep{Slowikowska2009}, \citet{Gonzalez2025} showed that a simple linear transformation of the optical Stokes parameters reproduces the phase-resolved polarization of IXPE, implying a common (likely synchrotron) origin. In this framework, the degree of X-ray polarization of pure pulsar emission mirrors the optical shape but is reduced by a factor $\sim 0.46$–$0.56$, while the polarization angle behavior exhibits a phase shift that is marginally consistent with zero at the MP in all three IXPE observations and variable at the IP. The near-zero MP lag points to an emission site far from the stellar surface, near or beyond the light cylinder, whereas the IP variability suggests a region subject to magnetospheric rearrangement. 

Following the glitches of July and August 2025, IXPE observed the Crab. We present here the polarimetric results of those observations. Despite the glitches being weaker than in 2019, the superior performances of IXPE provide us with more significant estimates on the amount of variation in PF or PA. We are then in a position to test the Polar-light findings, and to quantify the level of any compatible rearrangements in the pulsar magnetosphere. In the following discussion we have tried to maintain as much as possible consistency of notation/presentation with previous works to facilitate comparison, while providing an underlying unity of exposition.

\section{IXPE Observations} 
\label{sec:IXPE Observation}
Here we report polarimetric observations of the Crab Pulsar and Nebula following the two glitches on 17 July and 6 August 2025,  obtained with the Imaging X-ray Polarimetry Explorer (IXPE), a NASA mission in partnership with the Italian space agency (ASI). As described in detail elsewhere \citep{Weisskopf2022,Soffitta2021}, the IXPE Observatory includes three identical X-ray telescopes, each comprising an X-ray mirror assembly (NASA-furnished) and a polarization-sensitive pixelated detector \citep[][ASI-furnished]{Baldini2021}, to provide imaging polarimetry over a nominal 2-8 keV band. IXPE data telemetered to ground stations in Malindi (primary) and Singapore (secondary) are transmitted to the Mission Operations Center (MOC, at the Laboratory for Atmospheric and Space Physics, University of Colorado) and then to the Science Operations Center (SOC, at the NASA Marshall Space Flight Center). Using software developed jointly by ASI and NASA, the SOC processes science and relevant engineering and ancillary data, to produce data products that are archived at the High-Energy Astrophysics Science Archive Research Center (HEASARC, at the NASA Goddard Space Flight Center), for use by the international astrophysics community. We note that data from Detector Unit 2 were not available for this analysis due to an anomaly affecting its performance. 

\textit{IXPE} observed the Crab twice after the 2025 glitches: from 2025 September 9, 18{:}00~UT to September 15, 12{:}00~UT, and from 2025 October 1, 12{:}00~UT to October 3, 06{:}00~UT. Each campaign delivered a net exposure of $\approx 75$~ks. These observations are part of a regular monitoring program designed to increase the statistical significance of pulsar polarimetry. The first observing window was scheduled in response to the July and August 2025 glitches to probe the early post-glitch phase. The resulting start times place both datasets within post-glitch intervals of interest and, for the first window in particular, in a regime analogous to the first $\lesssim 100$~days after the 2019 glitch—when Polar-Light reported a $\sim$ 50\% reduced polarization fraction at an approximately constant position angle \citep{Feng2020,Long2021}
(see Tab. \ref{tab:ixpe_elapsed_start}). However we stress here that the 2019 glitch was much stronger than the 2025 ones, and that a comparison with the results  by PolarLight is meaningful only due to the superior performances of IXPE.

\begin{deluxetable*}{l l r}[ht!]
\tabletypesize{\footnotesize}
\tablecaption{Elapsed time (days) from the 2025 Crab glitches to the \textit{start} of each of the two successive IXPE observing window \label{tab:ixpe_elapsed_start}}
\tablehead{
\colhead{Glitch} & \colhead{IXPE observing window start (UTC)} & \colhead{$\Delta t_{\rm start}$ (days)}
}
\startdata
2025-07-17 (ATel~\#17298) & 2025-09-09 18{:}00 & 53.80 \\
2025-08-06 (ATel~\#17331) & 2025-09-09 18{:}00 & 34.86 \\
2025-07-17 (ATel~\#17298) & 2025-10-01 12{:}00 & 75.55 \\
2025-08-06 (ATel~\#17331) & 2025-10-01 12{:}00 & 56.62 \\
\enddata
\tablecomments{Glitch epochs adopted from ATel~\#17298 (MJD 60873.95; 2025-07-17 22{:}48 UT) and ATel~\#17331 (MJD 60893.9; 2025-08-06 21{:}36 UT). IXPE windows begin at 2025-09-09 18{:}00~UT and 2025-10-01 12{:}00~UT. $\Delta t_{\rm start}$ is the time between each glitch and the corresponding IXPE window start, in days.}
\end{deluxetable*}
\section{2025 Glitches Relative to the Full Dataset}
\begin{figure}[ht!]
\centering

\makebox[\textwidth][c]{%
  \includegraphics[width=0.54\textwidth]{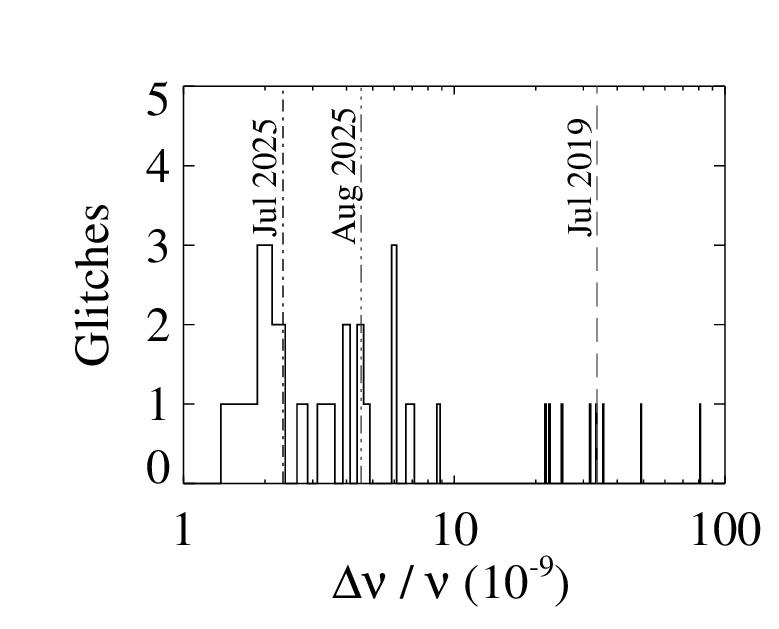}\hfill
  \includegraphics[width=0.54\textwidth]{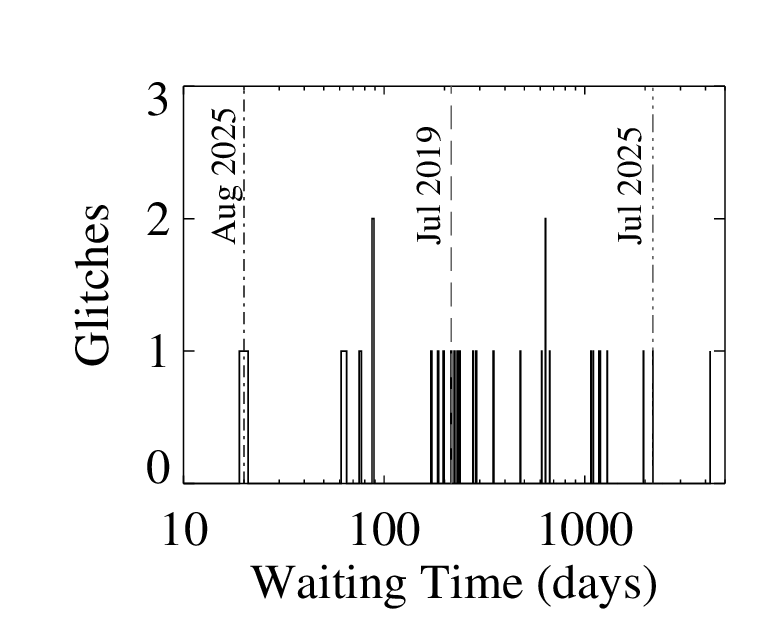}
}
\caption{({\bf a}) Histogram of Crab pulsar glitch sizes, shown as fractional frequency
  steps $\Delta\nu/\nu$ (log-scaled x–axis). Vertical markers indicate the
  glitches on 2019-07-23, 2025-07-17, and 2025-08-06. Historical events are compiled
  from the ATNF Pulsar Glitch Database for PSR~B0531+21/J0534+2200
  \citep{Manchester2005}, the 2019 event parameters are from \citet{Shaw2019},
  and the two 2025 glitches are from \cite{Shaw2025,Shaw2025b}. 
  ({\bf b}) Histogram of inter-glitch \emph{waiting times} for the Crab pulsar glitches.
  The x–axis is logarithmic and bins are thin to show the multi–decadal spread in intervals.
  Vertical markers label the waiting time immediately preceding three highlighted events:
  2019-07-23 ($\Delta t = 216.87\,\mathrm{d}$),
  2025-07-17 ($\Delta t = 2186.38\,\mathrm{d}$),
  and 2025-08-06 ($\Delta t = 19.95\,\mathrm{d}$).}
\label{fig:historical_glitches}
\end{figure}

The relative strengths of the 2019 glitch and the two 2025 glitches are shown in Fig.~\ref{fig:historical_glitches}, alongside the corresponding inter-glitch waiting times. Although the 2019 event exhibits a substantially larger fractional step in spin frequency than either of the 2025 glitches, the \emph{pair} of closely spaced 2025 events, separated by the shortest waiting times seen in the Crab and one of the longest waiting time for the July 2025 event, renders the IXPE observations especially valuable for probing the post-glitch magnetosphere. In particular, the doubled occurrence provides a natural effect for studying possible time-dependent changes in the magnetic-field geometry and polarization properties on dynamical timescales.

At present, no recovery timescale has been reported for the two Crab glitches of summer 2025. For reference (and context with \citet{Long2021}), following \citet{Shaw2021}, the delayed spin-up (rise) timescale and the main recovery (decay) post-glitch timescale after the 2019-07-23 event are $\tau_{\rm DSU}$ $\simeq$ 0.75 days and $\tau_{\rm rec}$$\simeq$ 6.4 days, respectively. The latter is compatible with a recovery of the pre-glitch polarization occurring after $\sim 50$--$100$~days, i.e. roughly $8$--$16$ e-folds of $\tau_{\rm rec}\approx 6.4$ days.

\section{Data Preparation for Phase-Resolved Analysis}
\label{sec:Data prep}

Given that the Crab ephemeris after the glitch was not available from Jodrell Bank at the time of writing, we used the sole X-ray photons to derive the post-glitch ephemeris for both Crab observations in the way 
reported in \citet{Bucciantini2023}. We first used as the starting point the Jodrell Bank monthly ephemeris of June 2025 in CGRO format\footnote{\url{https://www.jb.man.ac.uk/~pulsar/crab/CGRO_format.html}}, but modifying the frequency and derivatives for them to refer to an epoch between the September and early October 2025 observations.
The procedure requires first solving:
\begin{equation}
    \nu_{\text{new}} = \nu_{\text{old}} + \dot{\nu}_{\text{old}} (T_{\text{new}} - T_{\text{old}}) + 0.5 \ddot{\nu}_{\text{old}}(T_{\text{new}} - T_{\text{old}})^2
    \label{eq:velocity}
\end{equation}
\begin{equation}
    \dot{\nu}_{\text{new}} = \ddot{\nu}_{\text{old}} (T_{\text{new}} - T_{\text{old}})
    \label{eq:acceleration}
\end{equation}
where $\nu_{\mathrm{old}}$, $\dot{\nu}_{\mathrm{old}}$ and $\ddot{\nu}_{\mathrm{old}}$ are the frequency, its first and second time derivative at $T_{\mathrm{old}}$, while $\nu_{\mathrm{new}}$ and $\dot{\nu}_{\mathrm{new}}$ are the frequency and its first time derivative at $T_{\mathrm{new}}$.

This raw solution was, however, still inadequate for our purposes, due to the long time span from the July ephemeris and, of course, the glitches. We searched in a 0.002\,Hz-wide frequency interval around ${\nu}_{\text{new}}$ and a 5$\cdot10^{-12}$\,Hz/s interval around $\dot{\nu}_{\text{new}}$ using the \texttt{HENzsearch} tool distributed with \texttt{HENDRICS}\footnote{\url{https://hendrics.stingray.science}}
\citep{Bachetti2018}, with the \texttt{--fast} option for a quick 2d frequency-frequency derivative plane, using 4 harmonics to describe the profile. We then refined the search in a smaller frequency range using 10 harmonics.
In this way, we obtained a more precise guess that we refined manually with \texttt{HENphaseogram}, like in previous papers.
We then started the same iterative procedure used by \citet{Bucciantini2023}: with \texttt{HENphaseogram} we calculated, using the integrated profile as a template, pulsar times of arrivals (TOAs) in 100 small time intervals of the observation, that we fitted using the \texttt{pintk} graphical interface to \texttt{PINT}\footnote{\url{ https://nanograv-pint.readthedocs.io}} \citep{Luo2021}
obtaining a new spin-down solution and new TOAs. We then fed this new solution to \texttt{HENphaseogram}, where we calculated a more precise pulse profile to be used as a template for better TOAs.
The iterative process ended when the improvement
in the fitting through \texttt{PINT} was smaller than the uncertainties. The Crab ephemeris, that we have adopted in this work, are the following: \texttt{PEpoch=60941.643837880445}, \texttt{F0=29.54534816796599}, \texttt{F1=-3.6614720994413595E-10}.

\section{Result Standard \textit{Pulse On}/\textit{Pulse Off} Analysis.} 
\label{sec:IXPEStdAn}
\begin{figure}[ht!]
\centering
\includegraphics[scale=0.65]{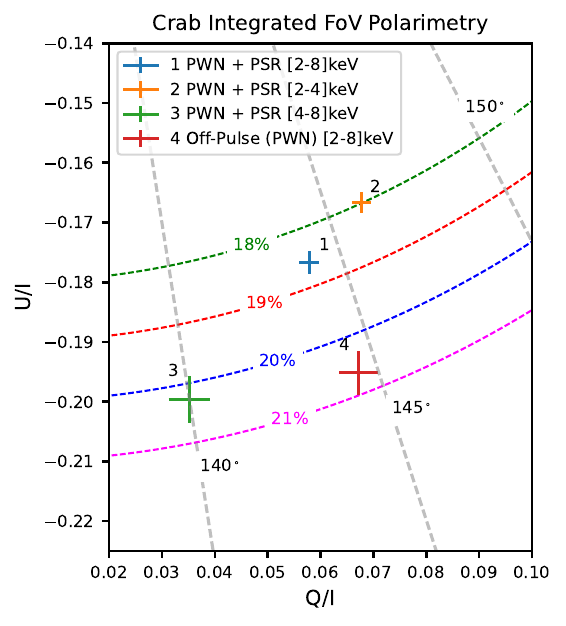}
\caption{Global polarization properties of the Crab pulsar and nebula (integrating all emission in a region within 2.5 arcmin of the PSR), in terms of normalized Stokes parameters for the electric field vector polarization angle (EVPA). The PF is expressed as percentage. Values have been obtained with the \texttt{PCUBE} algorithm
of \texttt{ixpeobssim}.}
\label{fig:high-lev1}
\end{figure}

\begin{figure}[ht!]
\centering
\includegraphics[scale=0.43]{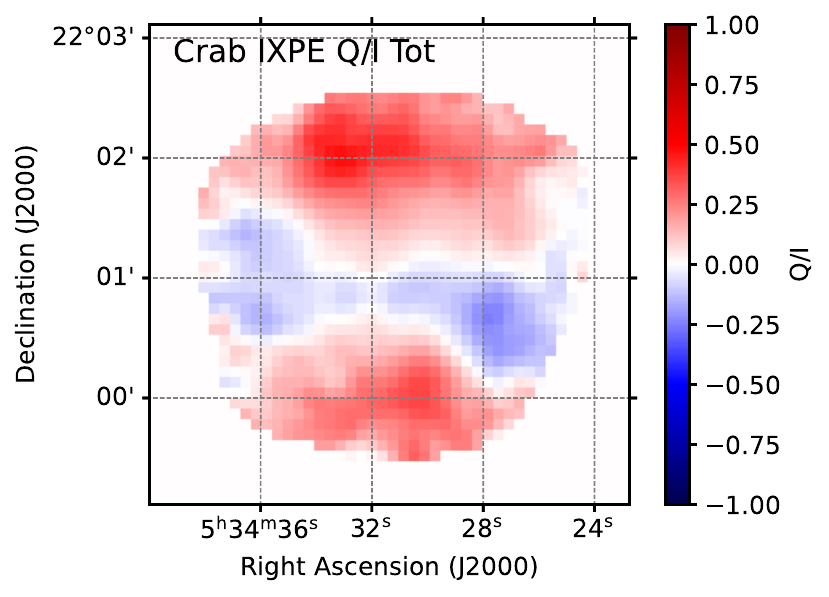}
\includegraphics[scale=0.43]{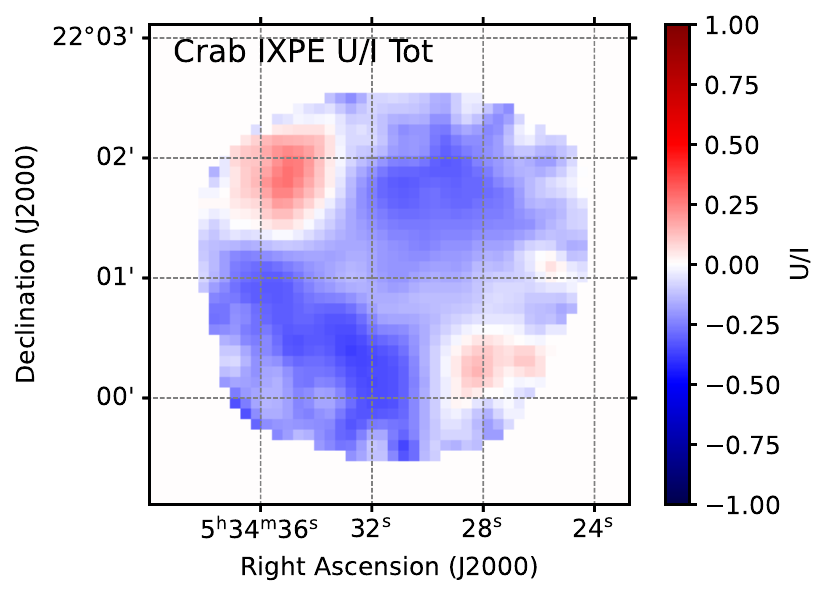}
\includegraphics[scale=0.43]{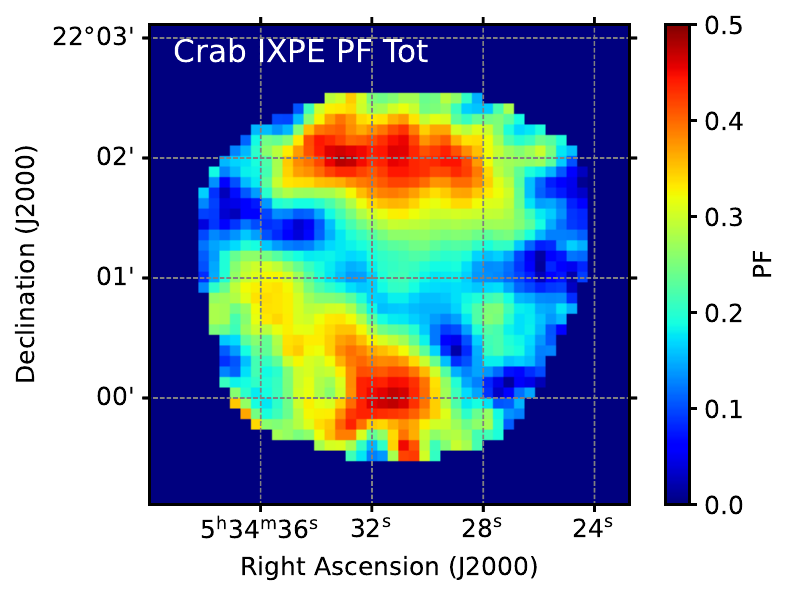}
\caption{Space resolved and phase integrated polarization properties of the Crab nebula and pulsar in the 2-8~keV band: left panel Q/I, central panel U/I and right panel PF. Data have been smoother with a Gaussian kernel (5~arcsec width), and limited to pixels having intensity above 0.5\% of the maximum. Maps have been obtained with the \texttt{PMAPCUBE} algorithm of \texttt{ixpeobssim}. See for comparison \citet{Bucciantini2023}.}
\label{fig:stokes_op}
\end{figure}

Here we present the results of a simple \textit{on-off analysis}, done following the exact same procedures of the first IXPE paper on the Crab nebula and pulsar by \citet{Bucciantini2023}, to which the reader is referred for further detailed information. Data reduction/processing was done using the \texttt{ixpeobssim}\footnote{\url{https://github.com/lucabaldini/ixpeobssim}} version 31.1.1 \citep{Baldini+22a}, adopting the latest \texttt{V013} Image Response Functions. In Fig.~\ref{fig:high-lev1} we 
 report the polarized properties of the Crab complex derived by  spatially integrating all emission in a region within 2.5~arcmin of the PSR. We recover the same variation in the PA between the low [2-4]~keV and high [4-8]~keV energy band observed previously, as well as the same change in
PF between the full and off-pulse (OP) emission. The OP polarization angle is fully consistent with previous results suggesting no significant change in the global polarization properties of the Crab nebula, since its first observation in 2022. \\
\\
In Tab.~\ref{tab:p1p2} we report the polarization properties measured in the main peak and in the interpulse. Phase ranges are the same of the first Crab paper for a direct comparison with those results. The  phase bin corresponding to center of the MP (here, as in the first paper, P1) has an OP subtracted emission  corresponding to PF = 15.4 ± 2.5\% and PA = 93$^\circ \pm 20^\circ$. This is fully consistent within 1$\sigma$ with previous findings. However we measure polarization at 3$\sigma$ confidence also in the left wing of the MP and at the center of the IP (here, as in the first paper, P2). The bridge appears to be unpolarized, with an upper limit at 3$\sigma$ of $\sim 30$\%. \\
\\
The nebula also displays the same raw polarization structure observed previously, as shown in Fig.~\ref{fig:stokes_op}. The degree of polarization has a clear north-south trend that does not match the position of the nebular axis. Localized regions of high PF $\sim 45\%$ can be seen at the very outer edge of the X-ray torus, while the central region shows a much lower level of polarization. The normalized Stokes parameters indicate a clear toroidal structure for the magnetic field geometry. We caution the reader that these results, obtained with the same approach as the one used in \citet{Bucciantini2023}, do not correct for polarization leakage \citep{Bucciantini+23b}, which however does not change the overall qualitative picture and contributes at most $\sim 10\%$ to the net PF in the outer regions of the torus.
\begin{deluxetable*}{l l l r}[ht!]
\tabletypesize{\footnotesize}
\tablecaption{Polarization properties of the main pulse (P1), interpulse (P2), Bridge and OP emission \label{tab:p1p2}}
\tablehead{
\colhead{Component} & \colhead{Q/I} & \colhead{U/I} & \colhead{PF}
}
\startdata
P1 left wing & $-0.107 \pm~0.034$ & $~~0.003 \pm~0.034$ & $0.108 \pm~0.034$\\
P1 center & $-0.106 \pm~0.027$ & $-0.012 \pm~0.027$ & $0.108 \pm~0.027$\\
P1 right wing & $~~0.032 \pm~0.041$ & $-0.008 \pm~0.041$ & $0.033 \pm~0.041$\\
P2 left wing & $-0.050 \pm~0.034$ & $~~0.002 \pm~0.034$& $0.050\pm~0.034$\\
P2 center & $-0.041 \pm~0.032$ &$-0.090\pm~0.032$ &$0.099\pm~0.032$\\
P2 right wing &$ -0.031 \pm~0.051$ &$ ~~0.044 \pm~0.051$ &$ 0.053\pm~0.051 $\\
Bridge & $~~0.006\pm 0.09$ & $~~0.008\pm 0.09$ & $0.010\pm 0.090$\\
OP & $-0.023\pm 0.009$ & $~~0.243\pm 0.009$ & $0.244\pm 0.009$\\
\enddata
\tablecomments{See \citet{Bucciantini2023} for a definition of the phase range of each component.}
\end{deluxetable*}
\\

\section{Results with Simultaneous Fitting for Nebula and Pulsar Polarization}
\label{sec:WongAnalyisis}

Here we present the results of applying the ``simultaneous fitting" technique of \cite{Wong2023} to measure the Crab pulsar and nebula polarization in the 2025 post-glitch IXPE observations. This method uses high temporal and spatial resolution observations provided by the Chandra X-ray Telescope and the IXPE instrumental response to simulate the pulsar and nebula components of the IXPE observations, which can be used to solve for their polarization via linear regression. This method yields higher sensitivity polarization measurements than the traditional On-Off method by replacing strict phase/spatial cuts with flux-based weights and can account for polarization leakage effects, although it does become model-dependent.

To generate the nebula model, we used the two Chandra observations from October \& December 2024 (Obs IDs 28052 and 30597, respectively) with a total exposure time of $\rm{\sim}\,10ks$. Despite not being coeval, no substantial difference in the nebular X-ray morphology was present between 2024 and 2022, suggesting a stable nebular structure. Readout streaks and the pulsar were removed using the method described in \cite{Wong2024}, and then the two observations were merged with the \texttt{CIAO} tool \texttt{reproject\_obs}. We apply this cleaned, merged Chandra observation to the most recent IXPE instrumental response (with validity date: 01 July 2024) to simulate the IXPE nebula observation using the \texttt{IXPEobssim} tool \texttt{xpobssim}. Pileup was corrected for using the method described in \cite{Wong2024}. 

To generate the pulsar model, we used the phase-resolved spectral measurements of the Crab Pulsar of \cite{Weisskopf2011} and similarly passed this through the IXPE instrumental response. For all simulations, an additional $6''$ Gaussian blur was applied to attain better agreement with the observed count maps, which may be attributed to calibration residuals in our PSF model. We also tried fitting with no additional blur, and the polarization measurements differ by $\lesssim0.1\sigma$ on average. The total simulation time is 1.5 Ms, about ${\sim}\,10\times$ IXPE exposure time, so that the models contribute negligibly to the final parameter errors.

\begin{figure}
\includegraphics[width=\linewidth]{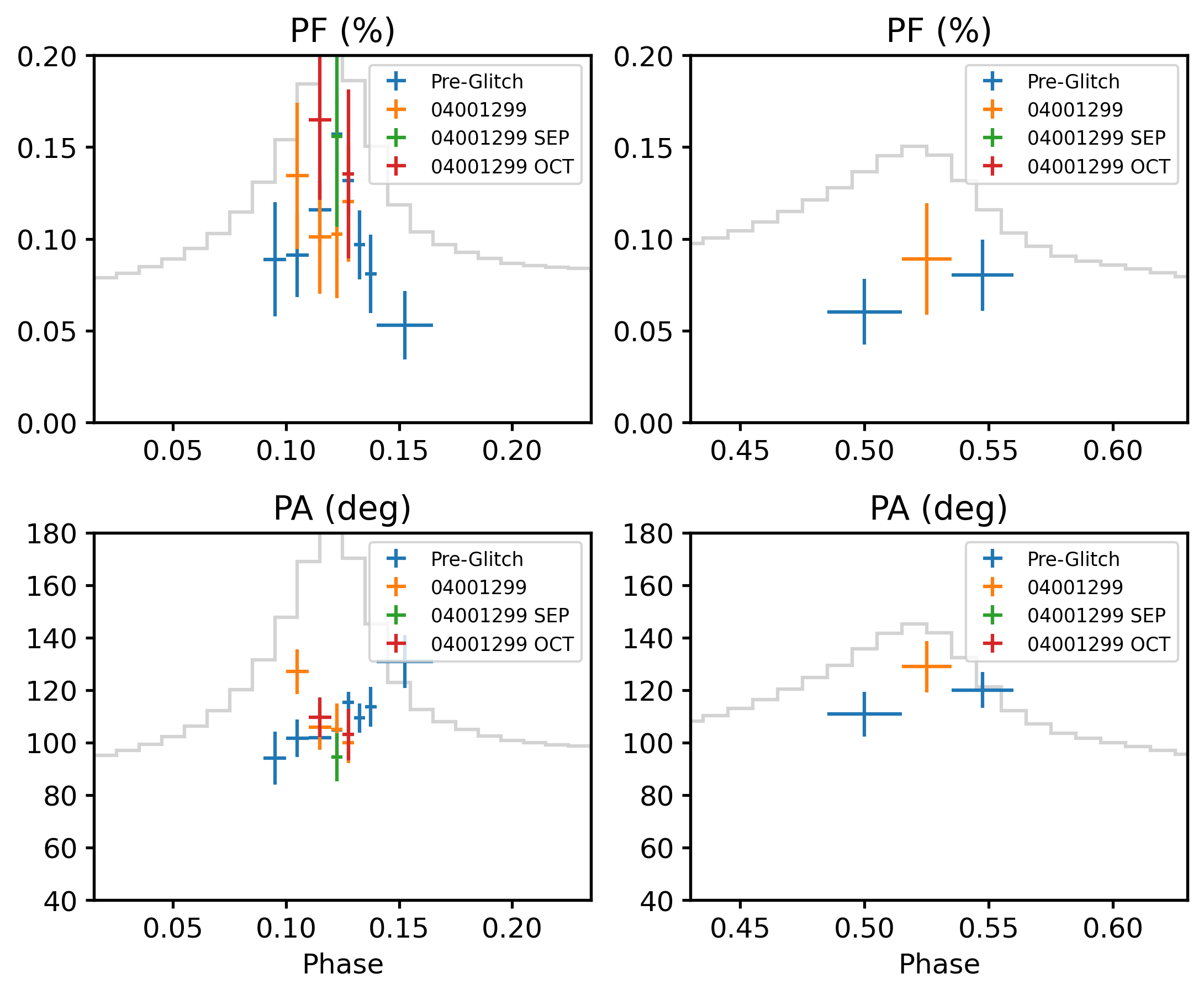}
\caption{Pre- and Post-Glitch Crab Pulsar Polarization obtained with Simultaneous Fitting. The method fits for the phase-varying pulsar and spatially-varying nebula polarization simultaneously using a known fixed flux model. Only ${>}\,2.8\sigma$ measurements are shown. No measurements at this significance level were found outside the Main Pulse (left) and Interpulse (right). Merged (orange) and individual Sep/Oct (green/red) post-glitch data are shown and are ${<}\,3\sigma$ from pre-glitch polarization at the same phase bin. Pre-glitch polarization were measured in \cite{Wong2024}. \label{fig:pulsar_pol}}
\end{figure}

\begin{figure}
\centering
\includegraphics[scale=0.15]{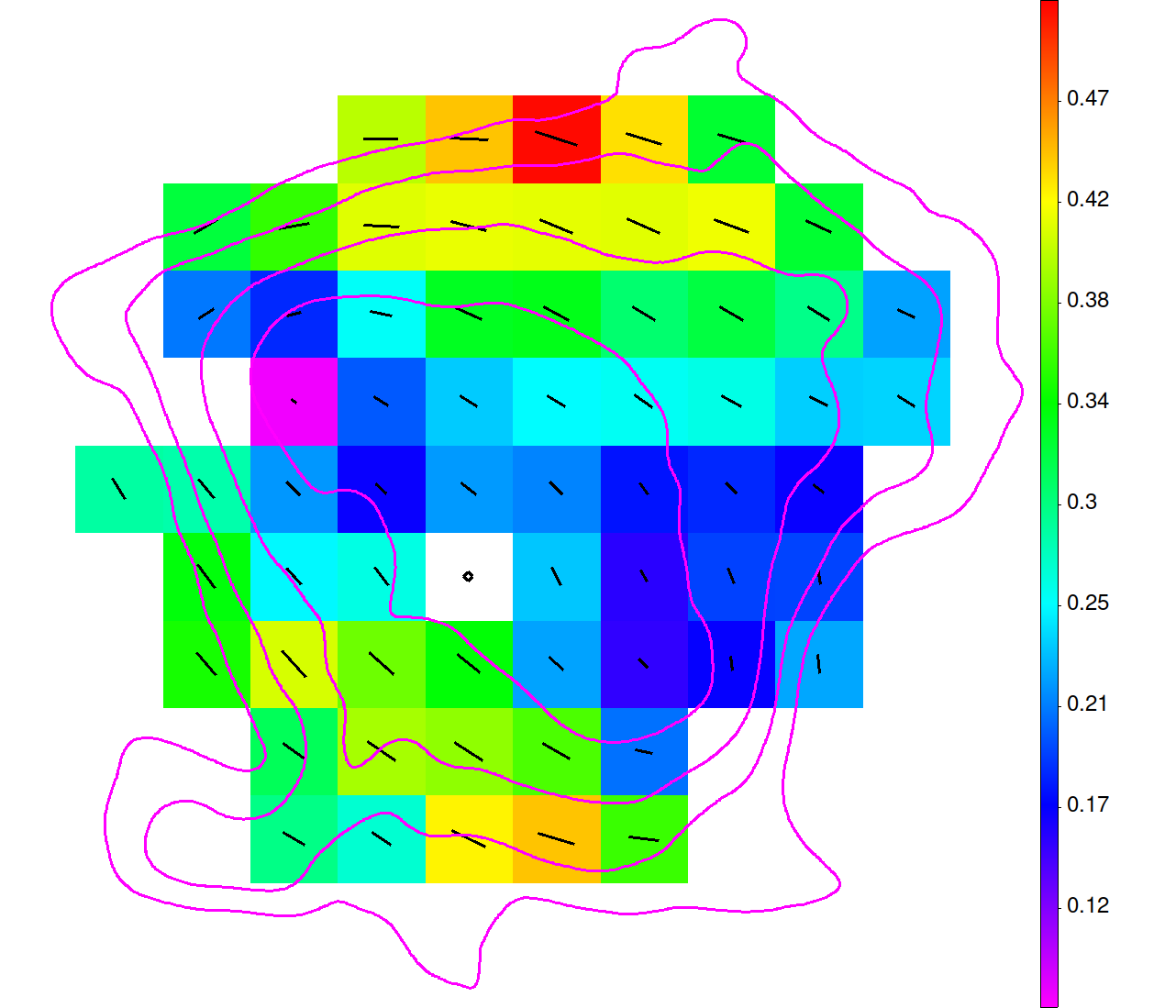}
\hspace{0.7cm}
\includegraphics[scale=0.15]{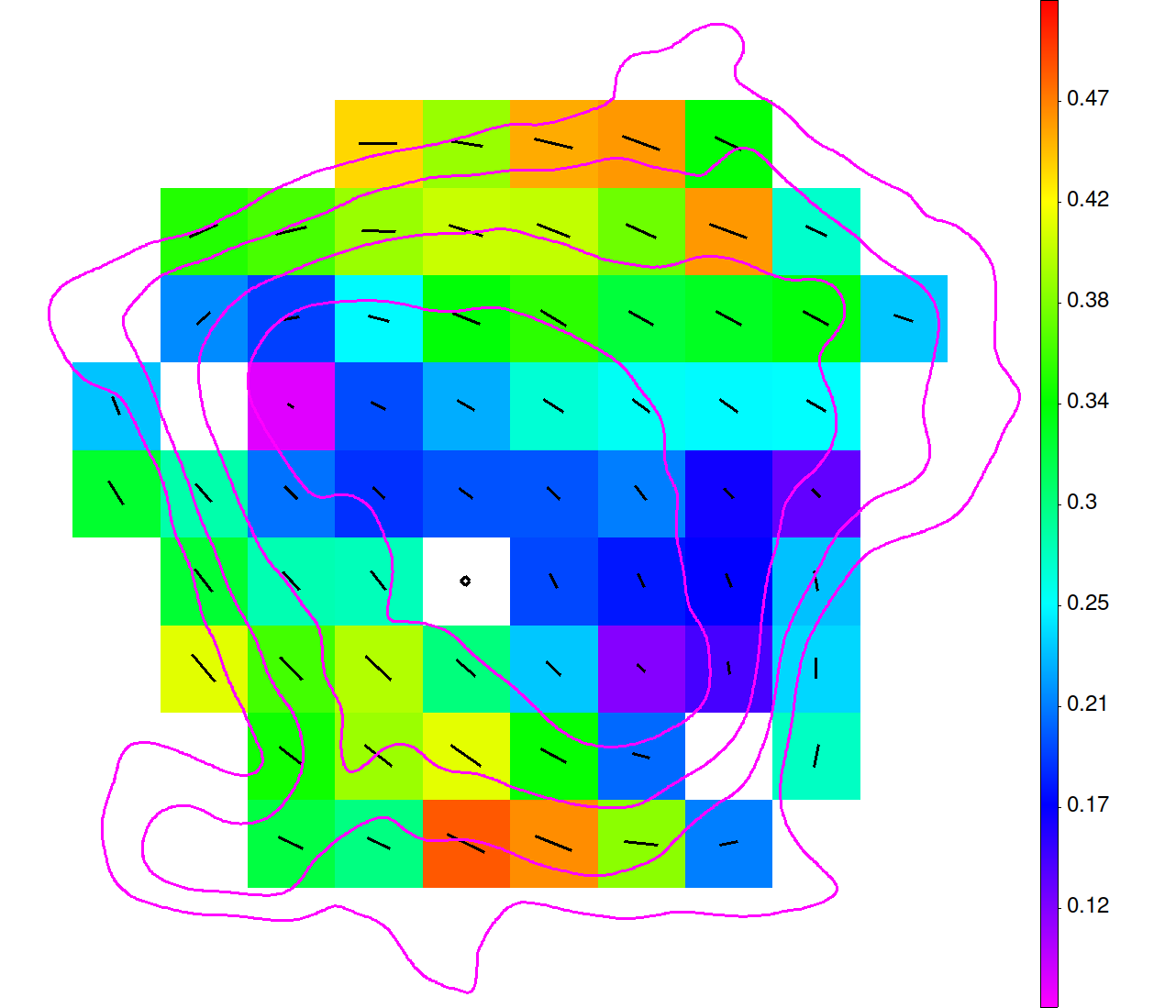}
\caption{ Sep/Oct (left/right) Post-Glitch Crab Nebula Polarization Maps obtained with Simultaneous Fitting. Matched colorbars range from PF = 8\% -- 51\%. Black vectors indicate magnetic field direction ($={\rm PA}+90^\circ$). Magenta Chandra contours are overlaid for reference. $5\sigma$ significance cut and 10,000-count flux cut have been applied. \label{fig:crab_neb}}
\end{figure}

We utilized the same binning scheme as in \cite{Wong2024}: $13 \times 13\ 15''$ pixels, one single 2--8 keV bin, and the same variable phase binning, with more smaller bins around the main and inter-pulses to map the rapid polarization change in these regions. The individual September and October post-glitch observations were analyzed separately as well as jointly. The pulsar phase-resolved polarization measurements are shown in Figure \ref{fig:pulsar_pol}. Figure \ref{fig:crab_neb} shows the nebula PF maps overlaid with magnetic field vectors for the two individual post-glitch observations. The integrated nebula polarization for the September and October observations are PF = $19.2 \pm 0.1\%$ and $19.1 \pm 0.1\%$ and PA = $145.0 \pm 0.15^\circ$ and $144.7 \pm 0.15^\circ$, respectively, and for the merged post-glitch observation is PF = $19.2 \pm 0.05\%$ and PA = $144.8 \pm 0.08^\circ$, consistent with the integrated nebular polarization reported in \cite{Bucciantini2023}, supporting the assumption that no major nebular structural change occurred.

No statistically significant differences in polarization between the pre- and post-glitch observations were found. In the nebula, the most polarized regions are located in the north/south, with PF ${\sim}\,45-50\%$, and the least polarized regions are located on the sides of the torus, where the polarization angle sweeps rapidly, with PF ${\sim}\,10-15\%$. These results are similar to those found in \cite{Bucciantini2023} and \cite{Wong2024}. For the pulsar, the greatest deviation between the merged pre-glitch and merged post-glitch observation lies in phase bin $0.1-0.11$. Hereafter, we study the upper-limit of the polarization difference between the merged pre and post-glitch observations at this phase bin. The difference in Stokes Q and U are 0.05 and 0.09, respectively. Considering the largest parameter uncertainty, $\rm Q_{err}\,{\sim}\,U_{err} = 0.04$ in the merged post-glitch observation, a 99\% upper limit for the difference in the Stokes parameters can be derived, yielding 0.15 and 0.20, respectively. The difference in PF and PA between the two observations are 4\% and $25^\circ$, and if we take the 99\% upper limit on the post-glitch polarization, 15\% and $41^\circ$. Now, we compare the Sept and merged post-glitch observations. For these observations, the only comparison that can be made is in phase bin $0.12-0.125$, where they both have polarization measurements that exceed ${>}\,2.8\sigma$. The difference in the Stokes parameters are $\rm\Delta Q =0.07 $ and $\rm\Delta U = 0.03$, with 99\% upper limits of 0.19 and 0.16, respectively. The difference in PF and PA are 5\% and $11^\circ$, with upper limits of 20\% and $155^\circ$. 

\section{Comparison of OPTIMA and IXPE phase resolved polarimetry.}
\label{sec:OptIXPEcomp}

\begin{figure}[ht!]
\begin{minipage}{0.42\columnwidth}
\includegraphics[width=\columnwidth]{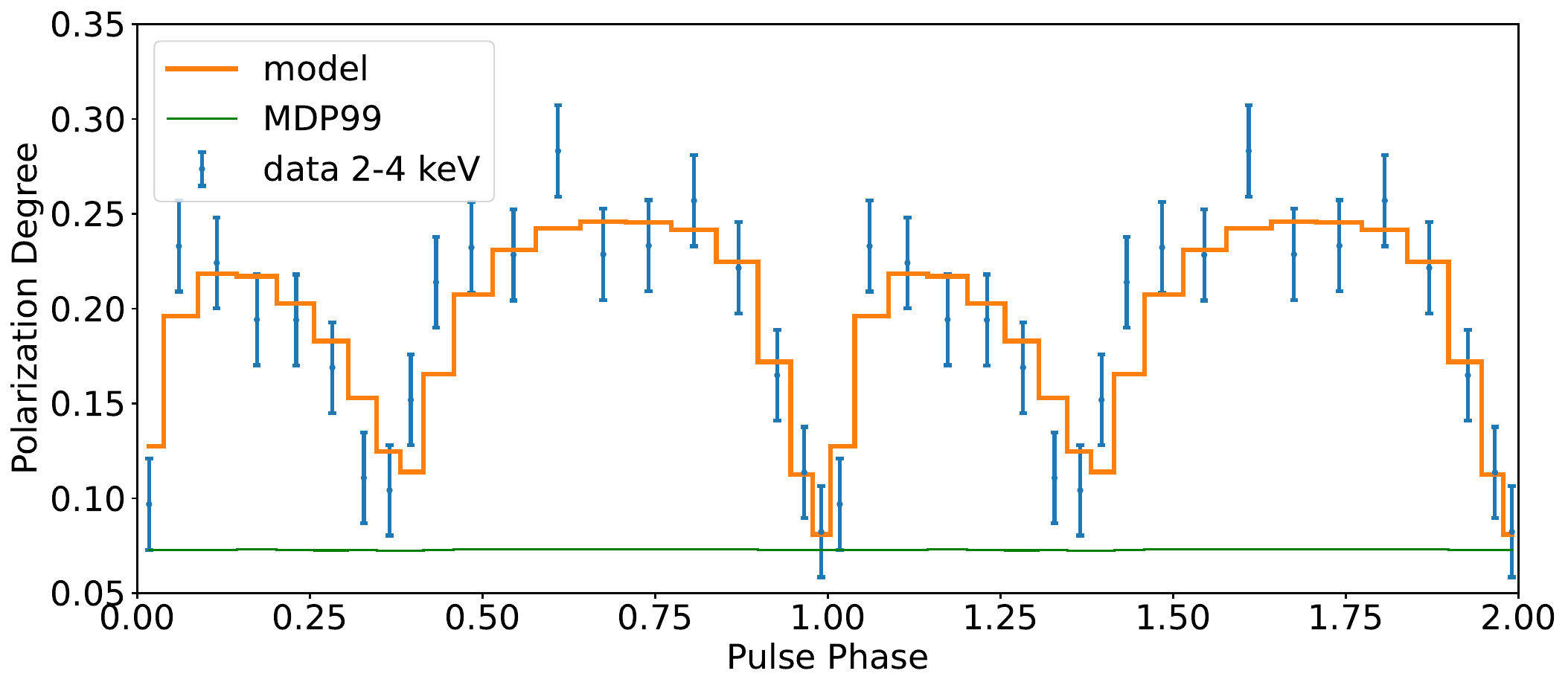}
\includegraphics[width=\columnwidth]{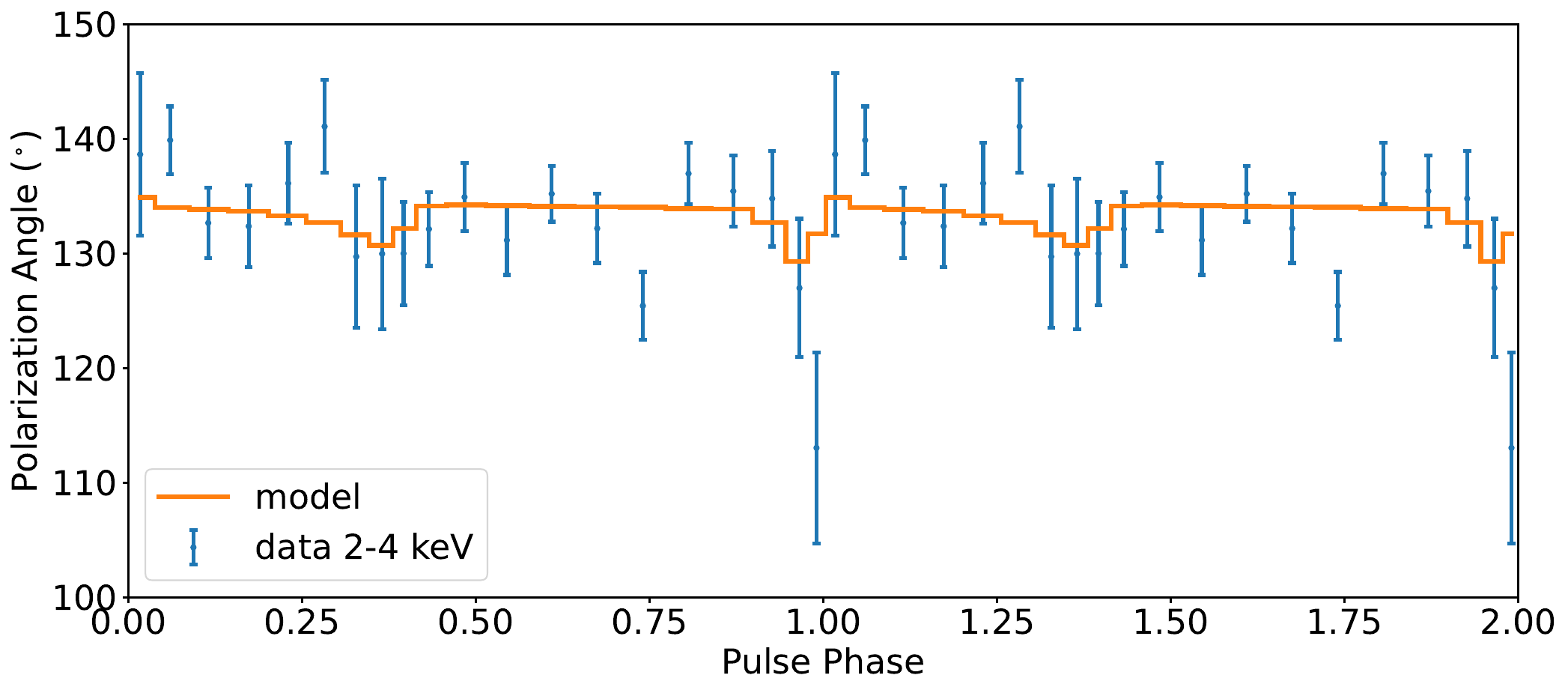}
\subfigure[\label{subfig:best_fit}]{\includegraphics[width=\columnwidth]{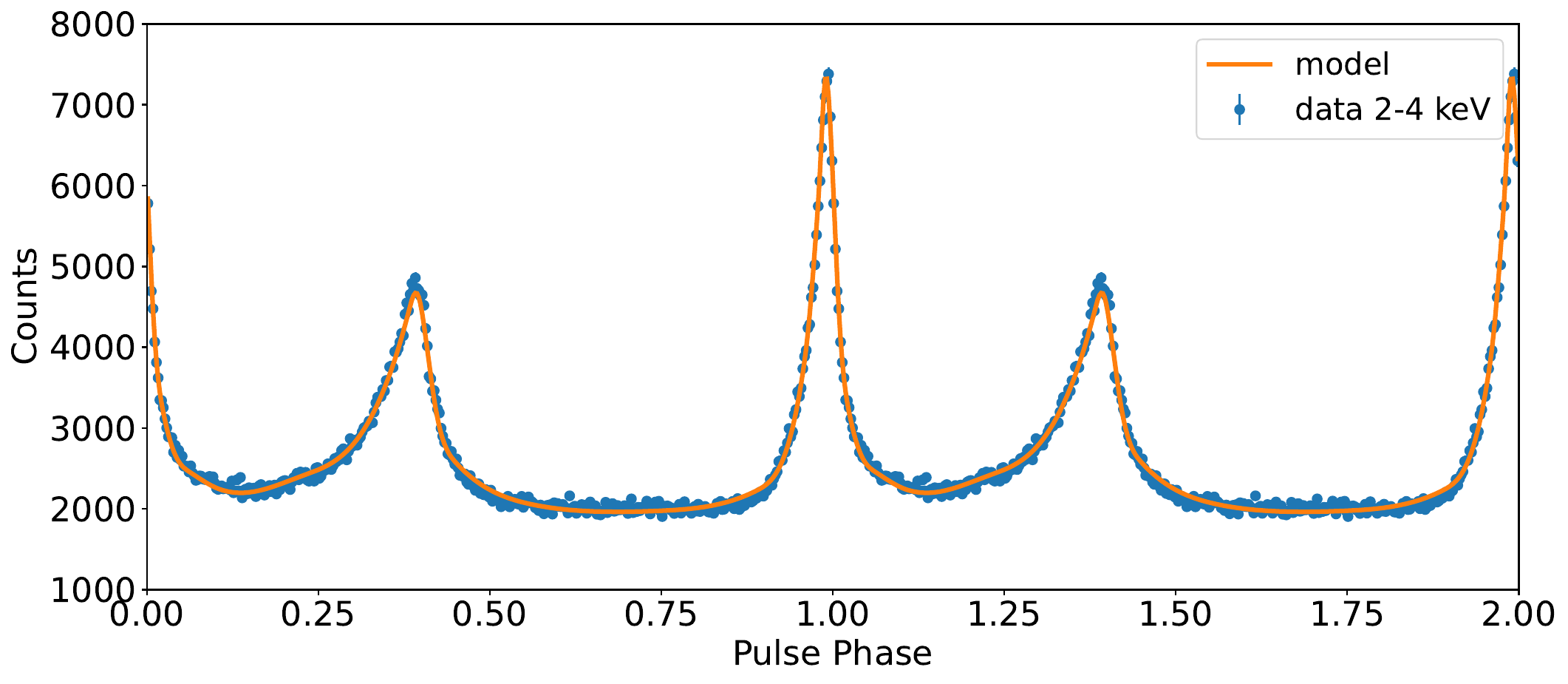}}
\end{minipage}
\begin{minipage}{0.57\columnwidth}
\subfigure[\label{subfig:corner}]{\includegraphics[width=\columnwidth]{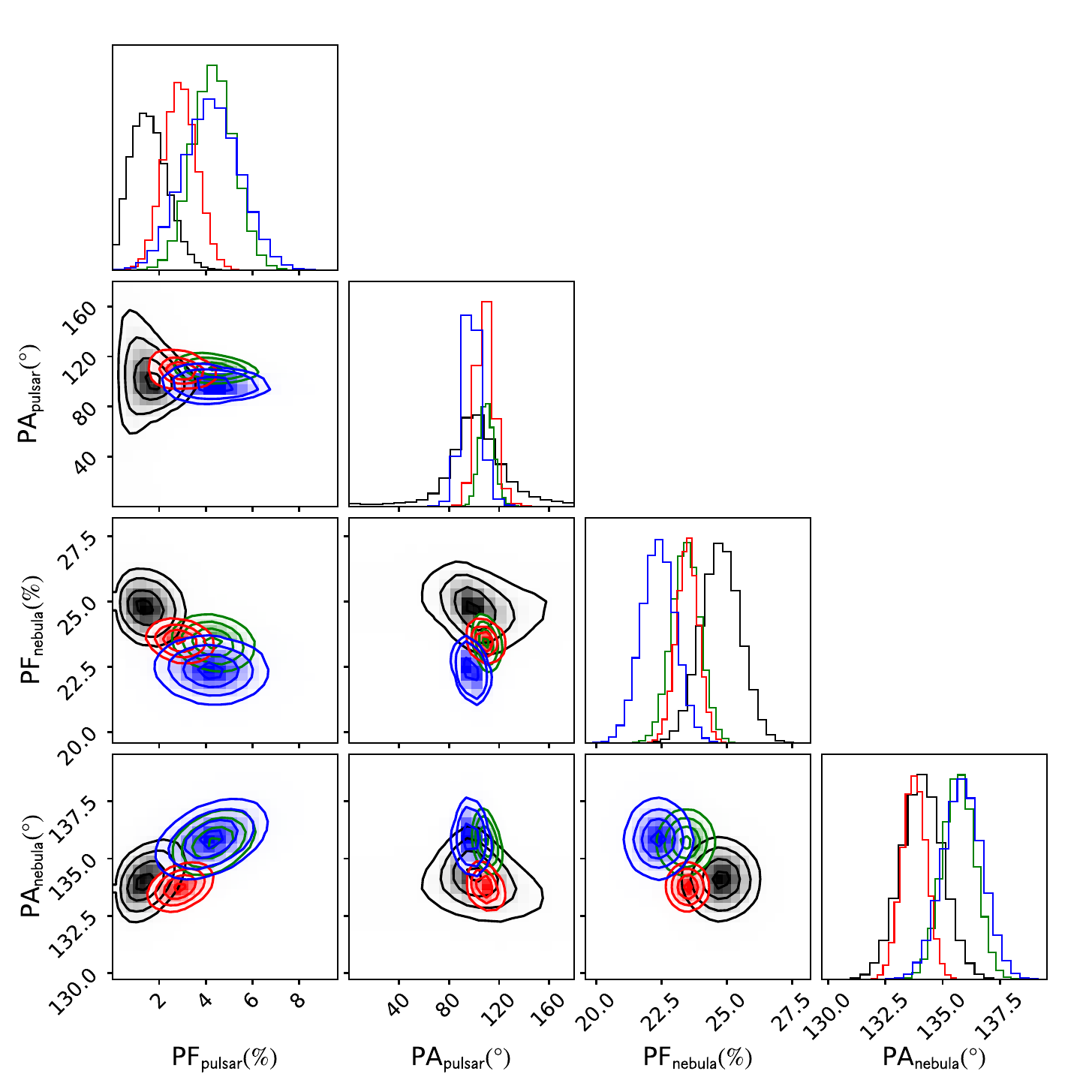}}
\end{minipage}
\caption{({\bf a}) Phase-dependent polarization properties of the Crab pulsar (two rotational cycles are shown for clarity). The blue data points correspond to the post-glitch (September) observation by IXPE (in the $2-4$ keV energy range), and the orange curve represents a phenomenological model based on phase-dependent optical (OPTIMA) and X-ray (Chandra) observations (see Sect. \ref{sec:OptIXPEcomp} for details). The top and middle panels show the polarization degree and polarization angle, respectively, computed using 20 equipopulated phase bins. The green line in the top panel indicates the minimum detectable polarization degree at the $99\%$ confidence level (MDP$_{99}$). The bottom panel displays the pulse profile using 400 equispaced phase bins. ({\bf b}) Corner plot showing the posterior distributions of the phase-averaged polarization parameters for the pure pulsar and pure nebula components derived from the phenomenological model using a MCMC analysis. The green, red, and blue contours correspond to the first, second, and third pre-glitch observations, respectively, while the black contours represent the September post-glitch observation.}
\end{figure}

As a further assessment of potential changes in the polarimetric properties following the glitch, we use the phenomenological model discussed in \cite{Gonzalez2025}. This model connects the phase-dependent polarization properties of the Crab pulsar between optical and X-rays via a linear transformation of the Stokes parameters:
\begin{equation}
\mathbf{S}_\mathrm{X} \equiv \mathbf{A} \cdot \mathbf{S}_\mathrm{V}  + \mathbf{B},     
\label{eq:liner_transformation}
\end{equation}
where
\begin{equation}
\mathbf{S}_\mathrm{X}=\begin{pmatrix}  I_\mathrm{X}\\ Q_\mathrm{X} \\ U_\mathrm{X}  \end{pmatrix},\;
\mathbf{A} =\begin{pmatrix} \alpha & 0 & 0\\0 & \beta & 0\\0 & 0 & \gamma \end{pmatrix},\; 
\mathbf{S}_\mathrm{V}=\begin{pmatrix} I_\mathrm{V}\\ Q_\mathrm{V} \\ U_\mathrm{V}  \end{pmatrix},\;
\mathbf{B} =\begin{pmatrix} b_I \\ b_Q \\ b_U\end{pmatrix}.
\label{eq:linear_transformation_components}
\end{equation}
Here, $\mathbf{S}_\mathrm{X}$ corresponds to the phase-dependent Stokes parameters measured by IXPE. $\mathbf{S}_\mathrm{V}$ corresponds to the phase-dependent Stokes measured in the optical band with OPTIMA \citep{Slowikowska2009}, $\mathbf{B}$ is a vector with constants that corresponds to the emission of the nebula, and $\mathbf{A}$ is a matrix that stretches or contracts the Stokes parameters in the optical band to match the ones measured by IXPE. By removing $90\%$ of the DC component from the optical Stokes parameters, this transformation simply represents a two-component model consisting of pure pulsar emission ($\mathbf{A}\cdot \mathbf{S}_\mathrm{V}$) plus nebula emission ($\mathbf{B}$). 

It is worth noting that the form of the $\mathbf{A}$ matrix is slightly different from that discussed in \cite{Gonzalez2025}, which has diagonal components $(\alpha,\beta,\beta)$. The repeated $\beta$ term in that formulation implicitly assumes that the polarization angle of the pure pulsar emission is identical in the optical and X-ray bands. Instead, this assumption can be relaxed by considering a diagonal composed of different terms $(\alpha,\beta,\gamma)$, which allows for deviations in the polarization angle between the two bands.

If we replace $I_\mathrm{V}$ by the intensity measured by Chandra $I_\mathrm{CX}$ \citep{Weisskopf2011}, which contains almost no nebula emission, the linear transformation can be expanded as follows:
\begin{eqnarray}
 I_\mathrm{X} &=&  \alpha\, I_\mathrm{CX} + b_I \label{eqn:Ix} \\
Q_\mathrm{X} &=& \beta\, p_\mathrm{V}\, 
\cos{\left(2\,\psi_\mathrm{V}\right)}\,  
I_\mathrm{CX} + b_Q  \label{eqn:Qx}\\
U_\mathrm{X} &=& \gamma\, p_\mathrm{V}\, \sin{\left(2\,\psi_\mathrm{V}\right)}\,   I_\mathrm{CX} + b_U . \label{eqn:Ux}
\label{eq:linear_transformation_expansion}
\end{eqnarray}
From this, it is possible to isolate the pulsar X-rays polarization properties and compute phase-averaged quantities:
\begin{equation}
\mathrm{PF}_\mathrm{pulsar} = \frac{\sqrt{\beta\,\langle  p_\mathrm{V}\, 
\cos{\left(2\,\psi_\mathrm{V}\right)}\,  I_\mathrm{CX}\rangle^2 + 
\gamma\, \langle  p_\mathrm{V}\, \sin{\left(2\,\psi_\mathrm{V}\right)}\,  
I_\mathrm{CX}\rangle^2}}{\alpha\,\langle I_\mathrm{CX} \rangle}
\label{eq:pulsar_PF}
\end{equation}
and
\begin{equation}
\mathrm{PA}_\mathrm{pulsar} = \frac{1}{2}\arctan\left(\frac{\gamma\, \langle  p_\mathrm{V}\, \sin{\left(2\,\psi_\mathrm{V}\right)}\,  
I_\mathrm{CX}\rangle}{\beta\, \langle  p_\mathrm{V}\, \cos{\left(2\,\psi_\mathrm{V}\right)}\,  
I_\mathrm{CX}\rangle}\right).
\label{eq:pulsar_PA}
\end{equation}
Similarly, it is possible to isolate the nebula contribution and compute:
\begin{equation}
\mathrm{PF}_\mathrm{nebula}=\frac{\sqrt{b_Q^2 + b_U^2}}{b_I}
\label{eq:nebula_PF}
\end{equation}
and 
\begin{equation}
\mathrm{PA}_\mathrm{nebula}=\frac{1}{2}\arctan\left(\frac{b_U}{b_Q}\right).
\label{eq:nebula_PA}
\end{equation}

As in previous sections, we proceed with the analysis of the  data using \texttt{ixpeobssim} package.  Following \cite{Gonzalez2025}, we consider a circular subtraction region with a radius of $20''$ around the pulsar location and we perform an unweighted analysis in the $2-4$ keV range, where IXPE has its maximum sensitivity. The data are divided in 20 equipopulated phase bins, and posterior distributions for the entire set of free parameters $\{\alpha, \beta, \gamma, b_I, b_Q, b_U\}$ are derived using an MCMC analysis with 100 walkers and 10000 iterations \citep{Foreman-Mackey2013}. For the sake of comparison, we present the results for the post-glitch observation together with pre-glitch observations (first, second, and third IXPE observations of the Crab). 

Fig.~\ref{subfig:best_fit} shows the phenomenological model fitted to the phase-dependent data of the September post-glitch observation. The model describes fairly well the data except at the main peak, where the observed polarization angle (middle panel) includes one data points deviating from the model.  Fig.~\ref{subfig:corner} shows a corner plot for the phase-averaged polarization properties of the pulsar and nebula, separately. The posterior distributions for the pre- and (September) post-glitch  measurements are consistent within 1-2 $\sigma$, indicating that no significant variations are found for the pulsar polarization or nebula polarization properties after the glitch.  A summary with the results of this analysis can be found in Table~\ref{tab:phase-averaged}.

We also studied potential phase shifts $(\delta_\mathrm{I}, \delta_\mathrm{P})$ in the phase-dependent polarization angle between the optical and X-ray bands for the interpulse and main pulse, respectively. To facilitate comparison with previous results from \cite{Gonzalez2025}, we adopt the $\mathbf{A}$ matrix with diagonal elements $(\alpha, \beta, \beta)$, implying that the (unshifted) phase-dependent polarization angle for pure pulsar emission is assumed to be identical in the optical and X-ray bands. For the MCMC analysis, we proceed with the same settings as explained above. For the September post-glitch data, we obtain $\delta_\mathrm{I}= - 1.15^{+4.27}_{-5.25}\%$ and $\delta_\mathrm{P}=-2.41^{+3.91}_{-4.01}\%$, which are consistent within $1\sigma$ with zero. A positive phase shift would indicate that the polarization angle in the optical band lags behind the X-ray band (see also \citealt{Heyl2000} for a discussion of propagation effects in the pulsar magnetosphere that can lead to the formation of such phase shifts). These results are also consistent with the values reported by \cite{Gonzalez2025} for the first, second, and third IXPE observations of the Crab pulsar. 

We also repeated the analysis considering the combined September-October post-glitch data. In this case, given the larger statistic, we use the phase-dependent data divided in 40 phase bins. A summary of the results is provided at the end of Table~\ref{tab:phase-averaged}.  The results are consistent with those of the previous analysis and again reveal no significant changes in the pulsar's properties after the glitch.

\begin{deluxetable*}{r c c c c r r r}[h!]
\tabletypesize{\footnotesize}
\tablecaption{Summary of the polarization properties obtained by fitting a phenomenological model to the IXPE data in the $2-4$ keV range. All quantities are derived from an MCMC analysis (see Sect.~\ref{sec:OptIXPEcomp} for details). }
\label{tab:phase-averaged}
\tablehead{\colhead{Obs.} &
\colhead{$\mathrm{PF}_\mathrm{pulsar}~(\%)$} & 
\colhead{$\mathrm{PA}_\mathrm{pulsar}~(^\circ)$} & 
\colhead{$\mathrm{PF}_\mathrm{nebula}~(\%)$} & 
\colhead{$\mathrm{PA}_\mathrm{nebula}~(^\circ)$} &
\colhead{$\delta_\mathrm{I}~(\%)$} &
\colhead{$\delta_\mathrm{P}~(\%)$} &
\colhead{$\chi^2/\mathrm{d.o.f}^{\mathrm{(a)}}$}}
\startdata
$1^\mathrm{st}$ pre-glitch & $4.31^{+0.90}_{-0.88}$ & $110.15^{+6.07}_{-5.60}$ & $23.42^{+0.55}_{-0.55}$ & $135.71^{+0.70}_{-0.70}$ & $2.91^{+1.36}_{-2.22}$ & $-0.54^{+0.45}_{-0.56}$ & 126.01/76\\
$2^\mathrm{nd}$ pre-glitch & $2.91^{+0.70}_{-0.70}$ & $108.59^{+7.23}_{-6.49}$ & $23.49^{+0.42}_{-0.41}$ & $133.76^{+0.53}_{-0.53}$ & $32.91^{+1.23}_{-1.51}$ & $-0.99^{+0.57}_{-0.65}$ & 98.38/76\\
$3^\mathrm{rd}$ pre-glitch & $4.25^{+1.20}_{-1.19}$ & $97.53^{+7.49}_{-6.89}$ & $22.39^{+0.64}_{-0.66}$ & $135.85^{+0.87}_{-0.87}$ & $-4.51^{+4.41}_{+7.17}$ & $-0.78^{+0.85}_{-1.09}$ & 97.17/76\\
Sep. post-glitch &$1.53^{+0.84}_{-0.73}$ & $101.49^{+21.54}_{-18.39}$ & $24.76^{+0.75}_{-0.75}$ & $134.10^{+0.90}_{-0.90}$ & $-1.15^{+4.27}_{-5.25}$ & $-2.41^{+3.91}_{-4.01}$ & 48.14/36\\
Sep./Oct. post-glitch &$2.30^{+0.62}_{-0.60}$ & $102.82^{+7.46}_{-6.76}$ & $24.12^{+0.53}_{-0.53}$ & $134.80^{+0.65}_{-0.65}$ & $0.22^{+2.36}_{-4.08}$ & $-1.01^{+0.75}_{-1.10}$ & 90.24/76  
\enddata
\tablecomments{(a) $\chi^2$ statistic reported for the fit to the Stokes parameters $Q$ and $U$.} 
\end{deluxetable*}

\section{Discussion}
\label{sec:Discussion}

In this section, we show how a measure of the changes in polarization properties can be translated into an assessment of possible magnetospheric variations. Adopting a simplified geometric model for the emission like the standard rotating vector model (RVM) we derive an expression useful for obtaining information on the maximum allowed change in magnetic obliquity by the upper limit change evaluated in the polarization angle after the glitch. 
All analyzes were performed by considering the September and October 2025 post-glitch observations both separately and jointly.

\begin{figure}[ht!]
    \centering
    \includegraphics[width=0.75\linewidth]{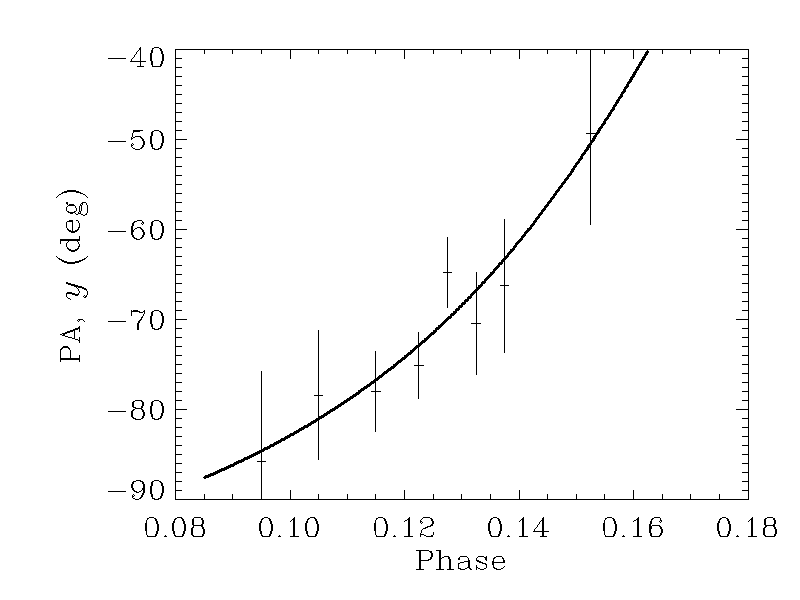}
    \caption{Crab Pulsar pre-glitch polarization angle and RVM fit.}
    \label{fig:RVM Fit}
\end{figure}

In the RVM it is possible to relate the observed  phase dependent polarization angle (PA = $\Psi$) to the  the viewing angle \textbf{$\zeta$}, and the phase $\phi$ as:
\begin{equation}
    \tan(\Psi(\phi) - \Psi_0) = \frac{\sin \alpha \sin (\phi - \phi_0)}{\sin \zeta \cos \alpha - \cos \zeta \sin \alpha \cos (\phi - \phi_0)}
    \label{eq:RVM}
\end{equation}
and its change before and after the glitch, to the  the viewing angle $\zeta$, and the phase $\phi$ as:
\begin{equation}
    \delta \Psi(\phi) \approx \frac{\partial \Psi}{\partial \alpha} \delta \alpha + \frac{\partial \Psi}{\partial \zeta} \delta \zeta + \frac{d \Psi}{d \phi} \delta \phi.
    \label{eq:placeholder_label}
\end{equation}
with: 
\begin{equation}
\frac{\partial \Psi}{\partial \alpha}(\phi)
=
\frac{1}{2}\,
\frac{\sin\zeta \, \sin(\phi-\phi_0)}
{\sin^2\alpha\,\sin^2(\phi-\phi_0)
+
\left[
\sin\zeta\,\cos\alpha
-
\cos\zeta\,\sin\alpha\,\cos(\phi-\phi_0)
\right]^2}
\label{eq:dpsi_dalpha}
\end{equation}
%
%
Thus, an upper limit $\delta\Psi_{max}$ in the PA change, due to any non detection, can be translated in a change in the magnetic obliquity $\alpha$ 
as:
\begin{equation}
    \label{eq:delapsi_psi}
    |\Delta \alpha| \leq \frac{\delta \Psi_{\text{max}}}{\left|\partial \Psi/\partial \alpha\right|}
\end{equation}
Data from the pre-glitch Crab observation are fitted to obtain the best-fit value of the parameter 
$\lvert \frac{\partial \Psi}{\partial \alpha} \rvert$ in Eq.~\ref{eq:dpsi_dalpha}, as shown in Fig. \ref{fig:RVM Fit}. The goodness-of-fit parameters are weakly constrained owing to the limited range of pulsar phase explored in the current analysis. The best-fit values were obtained by imposing the bounds $\alpha \in [1^\circ, 89^\circ]$ and $\zeta \in [1^\circ, 179^\circ]$. The corresponding fit results are summarized in Table~\ref{tab:RVM Fit}.

\begin{deluxetable*}{lcc}
\tablecaption{Parameters obtained from fitting the pre-glitch data with a rotating vector model.}\label{tab:RVM Fit}
\tablehead{
\colhead{Parameter} &
\colhead{Value ($^\circ$)} &
\colhead{Error (1$\sigma$) ($^\circ$)}
}
\startdata
$\alpha$ & 89.0 & \nodata \\
$\zeta$  & 91.8 & 4.0 \\
$\phi_0$ & 0.028 & 0.003 \\
$\Psi_0$ & $-26.7$ & 1.0 \\
\enddata
\end{deluxetable*}

As can be seen, $\alpha$ lies at the boundary of the allowed parameter interval, reflecting the weak constraints of the model imposed by the narrow pulsar phase range explored. The best-fit parameters were then inserted into Eq.~\ref{eq:dpsi_dalpha}, yielding a nearly phase-independent value of $ \lvert \partial \Psi / \partial \alpha (\phi)\rvert \simeq 5.0~\mathrm{rad\,rad^{-1}}$. This result translates into an upper limit on $\Delta\alpha$, reported in Table~\ref{tab:Dalpha}. 
In fact, when the parameter bounds are removed, the fit becomes essentially unconstrained. In this case, the best-fit parameters are $\alpha = 22.5^\circ$ and $\zeta = 14.4^\circ$, yielding $\lvert \partial \Psi / \partial \alpha (\phi) \rvert \simeq 10~\mathrm{rad\,rad^{-1}}$. This corresponds to a constraint on $\alpha$ that is approximately a factor of two tighter than that reported in Table~\ref{tab:Dalpha}.  
Alternatively, if $\zeta$ is fixed to $61.3^\circ$, corresponding to the angle of spin position in the sky derived by \cite{Ng2004}, we obtain $\alpha = 2.3^\circ$. In this case, the value of $\lvert \partial \Psi / \partial \alpha (\phi) \rvert$ is comparable to that obtained when the parameters are constrained, resulting in a limit on $\Delta\alpha$ consistent with the values reported in Table~\ref{tab:Dalpha}.

\begin{deluxetable*}{lc}[ht!]
\tablecaption{Maximum allowed variation of the magnetic inclination angle after the glitch}
\label{tab:Dalpha}
\tablehead{
\colhead{Data set} &
\colhead{$\Delta\alpha_{\mathrm{max}}$}
}
\startdata
Post-glitch (all) & $+4.0^\circ;\,-4.6^\circ$ \\
Post-glitch (Sep) & $+4.4^\circ;\,-7.2^\circ$ \\
Post-glitch (Oct) & $+5.0^\circ;\,-4.2^\circ$ \\
\enddata
\end{deluxetable*}

\section {Conclusions}
Polarization measures done by PolarLight  following a large Crab glitch in 2019, reported a significant change in the PSR polarization properties. Driven by these results, and taking advantage of its much higher polarization sensitivity, IXPE observed the Crab Nebula and pulsar, approximately 35–75 days after
the July and August 2025 glitches, twice, in early September and early October of the same year. This was done as part of a multi-epoch observing campaign aimed at deriving and modeling the phase-resolved polarization properties of the pulsar. Despite those glitches being weaker than the 2019 one, observed by PolarLight, we were able to set some tighter constraints on the change of polarization properties by applying several different data analysis techniques that the IXPE collaboration has developed during these years, specifically optimized for the study of the Crab PSR polarization signal.

We present here a detailed search for post-glitch variations in the polarimetric properties. The main conclusions of this study were:

\begin{itemize}
    \item Concerning the nebula, no differences in the space resolved polarimetry were found, not even for the off-pulse emission. Indeed, no glitch-induced effect is expected given the light-travel time from the pulsar to the nebula. However, the inner region of the Crab PWN is known to vary on timescales that can be as short as a few months. This also ensures that any possible variation of polarization in the pulsar region, if found, could safely be attributed to the PSR itself.
    \item Using a Pulse On/Pulse Off analysis, the polarization properties of the main pulse (MP) and interpulse (IP) were found to be unchanged within $1\sigma$, with respect to the pre-glitches IXPE results.
    \item A more sophisticated simultaneous angular and phase-resolved fit of Chandra and IXPE data for the nebula and pulsar, used to constrain the phase-resolved polarization properties of the pulsar, revealed no bin-to-bin variations within $3\sigma$.
    \item By comparing OPTIMA and \textit{IXPE} optical phase-resolved polarimetry, we searched for significant phase shifts between the two datasets and found no statistically significant changes.
\end{itemize}

Despite the fact that the latest glitches are among the weakest of the known population in the Crab, while the 2019 one was among the strongest, the excellent performance of IXPE allows us to set some stringent constraints  on associated effects on the magnetosphere. The two 2025 glitches had a minor effect on the magnetosphere, over medium to long terms. This  indicates the importance of a robust statistical dataset and analysis, to measure effects due to glitches. In the light of our results, even if the 2019 glitch was much stronger, previous claims of a very large depolarization of the pulsar signal, should be considered with caution, especially in the light of the lower statistics of previous measures.  We presented here an analysis carried out using three different approaches, complementary with each other. The basic on/off analysis of Sect.~\ref{sec:IXPEStdAn}, while totally agnostic for the nebula and pulsar properties, considers only events within a limited portion of the FOV (centered on the PSR and optimized to increase the ratio of the signal from PSR over the one from the PWN), and as such does not benefit from the full statistics of the observation (many events belonging to the PSR, but outside the region within 2.5 arcmin of the PSR, are lost). The simultaneous fitting of Sect.~\ref{sec:WongAnalyisis}, while totally agnostic for what concern the pulsar polarization properties, requires models for both the nebula and pulsar intensity, that are taken from Chandra data, but it is able to take full advantage of the entire statistic of the events over the FOV. Similarly the phase resolved polarimetric analysis of Sect.~\ref{sec:OptIXPEcomp} takes full advantage of the events statistics, but relies on prior knowledge of the pulsar’s optical polarization properties and X-ray intensity. 

We also attempted to interpret our results within the framework of a Rotating Vector Model applied to the available main-pulse X-ray Stokes parameters.  We derived a maximum change in pulsar obliquity allowed by the data of approximately $\pm 4^\circ$ at the 95\% confidence level. We remark however, that this finding is conditional on the adopted RVM geometry and parameter bounds, particularly given that the fit becomes
unconstrained when bounds are relaxed.

This represents the first study of its kind based on X-ray polarization data that combines phase-resolved and phase-averaged analyzes and includes a comparison with optical polarization measurements. Our results suggest that this approach provides an additional methodology for probing changes in the magnetospheric properties of the Crab pulsar, particularly in the case of glitches with much larger amplitudes or when using instruments with fast repointing capabilities and larger effective area.

\section{Software and third party data repository citations} \label{sec:cite}
   This paper employs a list of Chandra datasets, obtained by the Chandra X-ray Observatory, contained in the Chandra Data Collection (CDC) ~\dataset[DOI: 10.25574/cdc.553]{https://doi.org/10.25574/cdc.553}.
\section{Acknowledgments} 
The Imaging X-ray Polarimetry Explorer (IXPE) is a joint US and Italian mission.  The US contribution is supported by the National Aeronautics and Space Administration (NASA) and led and managed by its Marshall Space Flight Center (MSFC), with industry partner Ball Aerospace (now, BAE Systems).  The Italian contribution is supported by the Italian Space Agency (Agenzia Spaziale Italiana, ASI) through contract ASI-OHBI-2022-13-I.0, agreements ASI-INAF-2022-19-HH.0 and ASI-INFN-2017.13-H0, and its Space Science Data Center (SSDC) with agreements ASI-INAF-2022-14-HH.0 and ASI-INFN 2021-43-HH.0, and by the Istituto Nazionale di Astrofisica (INAF) and the Istituto Nazionale di Fisica Nucleare (INFN) in Italy. T.M. is supported by the JSPS KAKENHI Grant Numbers 23K25882 and 23H04895. J.W. is supported by NASA grants 80NSSC25K7523 and 80NSSC25K0277.  This research used data products provided by the IXPE Team (MSFC, SSDC, INAF, and INFN) and distributed with additional software tools by the High-Energy Astrophysics Science Archive Research Center (HEASARC), at NASA Goddard Space Flight Center (GSFC).

\bibliography{References}{}

@ARTICLE{Baldini+22a,
       author = {{Baldini}, Luca and {Bucciantini}, Niccol{\`o} and {Lalla}, Niccol{\`o} Di and {Ehlert}, Steven and {Manfreda}, Alberto and {Negro}, Michela and {Omodei}, Nicola and {Pesce-Rollins}, Melissa and {Sgr{\`o}}, Carmelo and {Silvestri}, Stefano},
        title = "{ixpeobssim: A simulation and analysis framework for the imaging X-ray polarimetry explorer}",
      journal = {SoftwareX},
         year = 2022,
        month = jul,
       volume = {19},
          eid = {101194},
        pages = {101194},
          doi = {10.1016/j.softx.2022.101194},
archivePrefix = {arXiv},
       eprint = {2203.06384},
 primaryClass = {astro-ph.IM},
       adsurl = {https://ui.adsabs.harvard.edu/abs/2022SoftX..1901194B}
}

@ARTICLE{Gugercinolu17,
       author = {{G{\"u}gercino{\u{g}}lu}, Erbil},
        title = "{Post-glitch exponential relaxation of radio pulsars and magnetars in terms of vortex creep across flux tubes}",
      journal = {\mnras},
         year = 2017,
        month = aug,
       volume = {469},
       number = {2},
        pages = {2313-2322},
          doi = {10.1093/mnras/stx985},
archivePrefix = {arXiv},
       eprint = {1701.05786},
 primaryClass = {astro-ph.HE},
       adsurl = {https://ui.adsabs.harvard.edu/abs/2017MNRAS.469.2313G}

}

@ARTICLE{Akbal+15b,
       author = {{Akbal}, O. and {G{\"u}gercino{\u{g}}lu}, E. and {{\c{S}}a{\c{s}}maz Mu{\c{s}}}, S. and {Alpar}, M.~A.},
        title = "{Peculiar glitch of PSR J1119-6127 and extension of the vortex creep model}",
      journal = {\mnras},
         year = 2015,
        month = may,
       volume = {449},
       number = {1},
        pages = {933-941},
          doi = {10.1093/mnras/stv322},
archivePrefix = {arXiv},
       eprint = {1502.03786},
 primaryClass = {astro-ph.HE},
       adsurl = {https://ui.adsabs.harvard.edu/abs/2015MNRAS.449..933A}
}

@ARTICLE{Antonopoulou+15a,
       author = {{Antonopoulou}, D. and {Weltevrede}, P. and {Espinoza}, C.~M. and {Watts}, A.~L. and {Johnston}, S. and {Shannon}, R.~M. and {Kerr}, M.},
        title = "{The unusual glitch recoveries of the high-magnetic-field pulsar J1119-6127}",
      journal = {\mnras},
         year = 2015,
        month = mar,
       volume = {447},
       number = {4},
        pages = {3924-3935},
          doi = {10.1093/mnras/stu2710},
archivePrefix = {arXiv},
       eprint = {1412.5853},
 primaryClass = {astro-ph.HE},
       adsurl = {https://ui.adsabs.harvard.edu/abs/2015MNRAS.447.3924A}
}

@ARTICLE{Bucciantini+23b,
       author = {{Bucciantini}, N. and {Di Lalla}, N. and {Romani}, R.~W.~R. and {Silvestri}, S. and {Negro}, M. and {Baldini}, L. and {Tennant}, A.~F. and {Manfreda}, A.},
        title = "{Polarisation leakage due to errors in track reconstruction in gas pixel detectors}",
      journal = {\aap},
         year = 2023,
        month = apr,
       volume = {672},
          eid = {A66},
        pages = {A66},
          doi = {10.1051/0004-6361/202245744},
archivePrefix = {arXiv},
       eprint = {2302.00346},
 primaryClass = {astro-ph.IM},
       adsurl = {https://ui.adsabs.harvard.edu/abs/2023A&A...672A..66B}
}

@ARTICLE{Alpar1984,
       author = {{Alpar}, M.~A. and {Pines}, D. and {Anderson}, P.~W. and {Shaham}, J.},
        title = "{Vortex creep and the internal temperature of neutron stars. I - General theory}",
      journal = {\apj},
         year = 1984,
        month = jan,
       volume = {276},
        pages = {325-334},
          doi = {10.1086/161616},
       adsurl = {https://ui.adsabs.harvard.edu/abs/1984ApJ...276..325A}
}

@ARTICLE{Anderson1975,
       author = {{Anderson}, P.~W. and {Itoh}, N.},
        title = "{Pulsar glitches and restlessness as a hard superfluidity phenomenon}",
      journal = {\nat},
         year = 1975,
        month = jul,
       volume = {256},
       number = {5512},
        pages = {25-27},
          doi = {10.1038/256025a0},
       adsurl = {https://ui.adsabs.harvard.edu/abs/1975Natur.256...25A}
}

@software{Bachetti2018,
       author = {{Bachetti}, Matteo},
        title = "{HENDRICS: High ENergy Data Reduction Interface from the Command Shell}",
 howpublished = {Astrophysics Source Code Library, record ascl:1805.019},
         year = 2018,
        month = may,
          eid = {ascl:1805.019},
archivePrefix = {ascl},
       eprint = {1805.019},
       adsurl = {https://ui.adsabs.harvard.edu/abs/2018ascl.soft05019B}
}

@ARTICLE{Baldini2021,
title = {Design, construction, and test of the Gas Pixel Detectors for the IXPE mission},
journal = {Astroparticle Physics},
volume = {133},
pages = {102628},
year = {2021},
issn = {0927-6505},
doi = {https://doi.org/10.1016/j.astropartphys.2021.102628},
url = {https://www.sciencedirect.com/science/article/pii/S0927650521000670},
author = {L. Baldini and M. Barbanera and R. Bellazzini and R. Bonino and F. Borotto and A. Brez and C. Caporale and C. Cardelli and S. Castellano and M. Ceccanti and S. Citraro and N. {Di Lalla} and L. Latronico and L. Lucchesi and C. Magazzu' and G. Magazzu' and S. Maldera and A. Manfreda and M. Marengo and A. Marrocchesi and P. Mereu and M. Minuti and F. Mosti and H. Nasimi and A. Nuti and C. Oppedisano and L. Orsini and M. Pesce-Rollins and M. Pinchera and A. Profeti and C. Sgro' and G. Spandre and M. Tardiola and D. Zanetti and F. Amici and H. Andersson and P. Attina' and M. Bachetti and W. Baumgartner and D. Brienza and R. Carpentiero and M. Castronuovo and L. Cavalli and E. Cavazzuti and M. Centrone and E. Costa and E. DAlba and F. D'Amico and E. {Del Monte} and S. {Di Cosimo} and A. {Di Marco} and G. {Di Persio} and I. Donnarumma and Y. Evangelista and S. Fabiani and R. Ferrazzoli and T. Kitaguchi and F. {La Monaca} and C. Lefevre and P. Loffredo and P. Lorenzi and E. Mangraviti and G. Matt and T. Meilahti and A. Morbidini and F. Muleri and T. Nakano and B. Negri and S. Nenonen and S.L. O'Dell and M. Perri and R. Piazzolla and S. Pieraccini and M. Pilia and S. Puccetti and B.D. Ramsey and J. Rankin and A. Ratheesh and A. Rubini and F. Santoli and P. Sarra and E. Scalise and A. Sciortino and P. Soffitta and T. Tamagawa and A.F. Tennant and A. Tobia and A. Trois and K. Uchiyama and M. Vimercati and M.C. Weisskopf and F. Xie and F. Zanetti and Y. Zhou}
}

@ARTICLE{Baym1969,
       author = {{Baym}, Gordon and {Pethick}, Christopher and {Pines}, David and {Ruderman}, Malvin},
        title = "{Spin Up in Neutron Stars : The Future of the Vela Pulsar}",
      journal = {\nat},
         year = 1969,
        month = nov,
       volume = {224},
       number = {5222},
        pages = {872-874},
          doi = {10.1038/224872a0},
       adsurl = {https://ui.adsabs.harvard.edu/abs/1969Natur.224..872B}
}

@ARTICLE{Beloborodov2009,
       author = {{Beloborodov}, Andrei M.},
        title = "{Untwisting Magnetospheres of Neutron Stars}",
      journal = {\apj},
         year = 2009,
        month = sep,
       volume = {703},
       number = {1},
        pages = {1044-1060},
          doi = {10.1088/0004-637X/703/1/1044},
archivePrefix = {arXiv},
       eprint = {0812.4873},
 primaryClass = {astro-ph},
       adsurl = {https://ui.adsabs.harvard.edu/abs/2009ApJ...703.1044B}
}

@ARTICLE{Bucciantini2023,
       author = {{Bucciantini}, Niccol{\`o} and {Ferrazzoli}, Riccardo and {Bachetti}, Matteo and {Rankin}, John and {Di Lalla}, Niccol{\`o} and {Sgr{\`o}}, Carmelo and {Omodei}, Nicola and {Kitaguchi}, Takao and {Mizuno}, Tsunefumi and {Gunji}, Shuichi and {Watanabe}, Eri and {Baldini}, Luca and {Slane}, Patrick and {Weisskopf}, Martin C. and {Romani}, Roger W. and {Possenti}, Andrea and {Marshall}, Herman L. and {Silvestri}, Stefano and {Pacciani}, Luigi and {Negro}, Michela and {Muleri}, Fabio and {de O{\~n}a Wilhelmi}, Emma and {Xie}, Fei and {Heyl}, Jeremy and {Pesce-Rollins}, Melissa and {Wong}, Josephine and {Pilia}, Maura and {Agudo}, Iv{\'a}n and {Antonelli}, Lucio A. and {Baumgartner}, Wayne H. and {Bellazzini}, Ronaldo and {Bianchi}, Stefano and {Bongiorno}, Stephen D. and {Bonino}, Raffaella and {Brez}, Alessandro and {Capitanio}, Fiamma and {Castellano}, Simone and {Cavazzuti}, Elisabetta and {Chen}, Chien-Ting and {Ciprini}, Stefano and {Costa}, Enrico and {De Rosa}, Alessandra and {Del Monte}, Ettore and {Di Gesu}, Laura and {Di Marco}, Alessandro and {Donnarumma}, Immacolata and {Doroshenko}, Victor and {Dov{\v{c}}iak}, Michal and {Ehlert}, Steven R. and {Enoto}, Teruaki and {Evangelista}, Yuri and {Fabiani}, Sergio and {Garcia}, Javier A. and {Hayashida}, Kiyoshi and {Iwakiri}, Wataru and {Jorstad}, Svetlana G. and {Kaaret}, Philip and {Karas}, Vladimir and {Kislat}, Fabian and {Kolodziejczak}, Jeffery J. and {Krawczynski}, Henric and {La Monaca}, Fabio and {Latronico}, Luca and {Liodakis}, Ioannis and {Maldera}, Simone and {Manfreda}, Alberto and {Marin}, Fr{\'e}d{\'e}ric and {Marinucci}, Andrea and {Marscher}, Alan P. and {Massaro}, Francesco and {Matt}, Giorgio and {Mitsuishi}, Ikuyuki and {Ng}, C. -Y. and {O'Dell}, Stephen L. and {Oppedisano}, Chiara and {Papitto}, Alessandro and {Pavlov}, George G. and {Peirson}, Abel L. and {Perri}, Matteo and {Petrucci}, Pierre-Olivier and {Poutanen}, Juri and {Puccetti}, Simonetta and {Ramsey}, Brian D. and {Ratheesh}, Ajay and {Roberts}, Oliver J. and {Soffitta}, Paolo and {Spandre}, Gloria and {Swartz}, Doug and {Tamagawa}, Toru and {Tavecchio}, Fabrizio and {Taverna}, Roberto and {Tawara}, Yuzuru and {Tennant}, Allyn F. and {Thomas}, Nicolas E. and {Tombesi}, Francesco and {Trois}, Alessio and {Tsygankov}, Sergey and {Turolla}, Roberto and {Vink}, Jacco and {Wu}, Kinwah and {Zane}, Silvia},
        title = "{Simultaneous space and phase resolved X-ray polarimetry of the Crab pulsar and nebula}",
      journal = {Nature Astronomy},
         year = 2023,
        month = may,
       volume = {7},
        pages = {602-610},
          doi = {10.1038/s41550-023-01936-8},
archivePrefix = {arXiv},
       eprint = {2207.05573},
 primaryClass = {astro-ph.HE},
       adsurl = {https://ui.adsabs.harvard.edu/abs/2023NatAs...7..602B}
}

@ARTICLE{Fabiani2014,
   author = {{Fabiani}, S. and {Costa}, E. and {Del Monte}, E. and {Muleri}, F. and
    {Soffitta}, P. and {Rubini}, A. and {Bellazzini}, R. and {Brez}, A. and
    {de Ruvo}, L. and {Minuti}, M. and {Pinchera}, M. and {Sgr{\'o}}, C. and
    {Spandre}, G. and {Spiga}, D. and {Tagliaferri}, G. and {Pareschi}, G. and
    {Basso}, S. and {Citterio}, O. and {Burwitz}, V. and {Burkert}, W. and
    {Menz}, B. and {Hartner}, G.},
    title = "{The Imaging Properties of the Gas Pixel Detector as a Focal Plane Polarimeter}",
  journal = {\apjs},
archivePrefix = "arXiv",
   eprint = {1403.7200},
 primaryClass = "astro-ph.IM",
     year = 2014,
    month = jun,
   volume = 212,
      eid = {25},
    pages = {25},
      doi = {10.1088/0067-0049/212/2/25},
   adsurl = {http://adsabs.harvard.edu/abs/2014ApJS..212...25F}
}

@ARTICLE{Feng2019,
       author = {{Feng}, Hua and {Jiang}, Weichun and {Minuti}, Massimo and {Wu}, Qiong and {Jung}, Aera and {Yang}, Dongxin and {Citraro}, Saverio and {Nasimi}, Hikmat and {Yu}, Jiandong and {Jin}, Ge and {Huang}, Jiahui and {Zeng}, Ming and {An}, Peng and {Baldini}, Luca and {Bellazzini}, Ronaldo and {Brez}, Alessandro and {Latronico}, Luca and {Sgr{\`o}}, Carmelo and {Spandre}, Gloria and {Pinchera}, Michele and {Muleri}, Fabio and {Soffitta}, Paolo and {Costa}, Enrico},
        title = "{PolarLight: a CubeSat X-ray polarimeter based on the gas pixel detector}",
      journal = {Experimental Astronomy},
         year = 2019,
        month = apr,
       volume = {47},
       number = {1-2},
        pages = {225-243},
          doi = {10.1007/s10686-019-09625-z},
archivePrefix = {arXiv},
       eprint = {1903.01619},
 primaryClass = {astro-ph.IM},
       adsurl = {https://ui.adsabs.harvard.edu/abs/2019ExA....47..225F}
}

@ARTICLE{Feng2020b,
       author = {{Feng}, Hua and {Li}, Hong and {Long}, Xiangyun and {Bellazzini}, Ronaldo and {Costa}, Enrico and {Wu}, Qiong and {Huang}, Jiahui and {Jiang}, Weichun and {Minuti}, Massimo and {Wang}, Weihua and {Xu}, Renxin and {Yang}, Dongxin and {Baldini}, Luca and {Citraro}, Saverio and {Nasimi}, Hikmat and {Soffitta}, Paolo and {Muleri}, Fabio and {Jung}, Aera and {Yu}, Jiandong and {Jin}, Ge and {Zeng}, Ming and {An}, Peng and {Brez}, Alessandro and {Latronico}, Luca and {Sgro}, Carmelo and {Spandre}, Gloria and {Pinchera}, Michele},
        title = "{Re-detection and a possible time variation of soft X-ray polarization from the Crab}",
      journal = {Nature Astronomy},
         year = 2020,
        month = may,
       volume = {4},
        pages = {511-516},
          doi = {10.1038/s41550-020-1088-1},
archivePrefix = {arXiv},
       eprint = {2011.05487},
 primaryClass = {astro-ph.HE},
       adsurl = {https://ui.adsabs.harvard.edu/abs/2020NatAs...4..511F}
}

@ARTICLE{Feng2020,
       author = {{Feng}, Hua and {Li}, Hong and {Long}, Xiangyun and {Bellazzini}, Ronaldo and {Costa}, Enrico and {Wu}, Qiong and {Huang}, Jiahui and {Jiang}, Weichun and {Minuti}, Massimo and {Wang}, Weihua and {Xu}, Renxin and {Yang}, Dongxin and {Baldini}, Luca and {Citraro}, Saverio and {Nasimi}, Hikmat and {Soffitta}, Paolo and {Muleri}, Fabio and {Jung}, Aera and {Yu}, Jiandong and {Jin}, Ge and {Zeng}, Ming and {An}, Peng and {Brez}, Alessandro and {Latronico}, Luca and {Sgro}, Carmelo and {Spandre}, Gloria and {Pinchera}, Michele},
        title = "{Author Correction: Re-detection and a possible time variation of soft X-ray polarization from the Crab}",
      journal = {Nature Astronomy},
         year = 2020,
        month = jun,
       volume = {4},
        pages = {808-808},
          doi = {10.1038/s41550-020-1159-3},
       adsurl = {https://ui.adsabs.harvard.edu/abs/2020NatAs...4..808F}
}

@ARTICLE{Foreman-Mackey2013,
       author = {{Foreman-Mackey}, Daniel and {Hogg}, David W. and {Lang}, Dustin and {Goodman}, Jonathan},
        title = "{emcee: The MCMC Hammer}",
      journal = {\pasp},
         year = 2013,
        month = mar,
       volume = {125},
       number = {925},
        pages = {306},
          doi = {10.1086/670067},
archivePrefix = {arXiv},
       eprint = {1202.3665},
 primaryClass = {astro-ph.IM},
       adsurl = {https://ui.adsabs.harvard.edu/abs/2013PASP..125..306F}
}

@ARTICLE{Fuentes2017,
       author = {{Fuentes}, J.~R. and {Espinoza}, C.~M. and {Reisenegger}, A. and {Shaw}, B. and {Stappers}, B.~W. and {Lyne}, A.~G.},
        title = "{The glitch activity of neutron stars}",
      journal = {\aap},
         year = 2017,
        month = dec,
       volume = {608},
          eid = {A131},
        pages = {A131},
          doi = {10.1051/0004-6361/201731519},
archivePrefix = {arXiv},
       eprint = {1710.00952},
 primaryClass = {astro-ph.HE},
       adsurl = {https://ui.adsabs.harvard.edu/abs/2017A&A...608A.131F}
}

@ARTICLE{Haskell2015,
       author = {{Haskell}, Brynmor and {Melatos}, Andrew},
        title = "{Models of pulsar glitches}",
      journal = {International Journal of Modern Physics D},
         year = 2015,
        month = jan,
       volume = {24},
       number = {3},
          eid = {1530008},
        pages = {1530008},
          doi = {10.1142/S0218271815300086},
archivePrefix = {arXiv},
       eprint = {1502.07062},
 primaryClass = {astro-ph.SR},
       adsurl = {https://ui.adsabs.harvard.edu/abs/2015IJMPD..2430008H}
}

@ARTICLE{Heyl2000,
   author = {{Heyl}, J.~S. and {Shaviv}, N.~J.},
    title = "{Polarization evolution in strong magnetic fields}",
  journal = {\mnras},
   eprint = {arXiv:astro-ph/9909339},
     year = 2000,
    month = jan,
   volume = 311,
    pages = {555-564},
      doi = {10.1046/j.1365-8711.2000.03076.x},
   adsurl = {http://adsabs.harvard.edu/abs/2000MNRAS.311..555H}
}

@ARTICLE{Kaspi2017,
       author = {{Kaspi}, Victoria M. and {Beloborodov}, Andrei M.},
        title = "{Magnetars}",
      journal = {\araa},
         year = 2017,
        month = aug,
       volume = {55},
       number = {1},
        pages = {261-301},
          doi = {10.1146/annurev-astro-081915-023329},
archivePrefix = {arXiv},
       eprint = {1703.00068},
 primaryClass = {astro-ph.HE},
       adsurl = {https://ui.adsabs.harvard.edu/abs/2017ARA&A..55..261K}
}

@ARTICLE{Long2021,
       author = {{Long}, Xiangyun and {Feng}, Hua and {Li}, Hong and {Zhu}, Jiahuan and {Wu}, Qiong and {Huang}, Jiahui and {Minuti}, Massimo and {Jiang}, Weichun and {Wang}, Weihua and {Xu}, Renxin and {Costa}, Enrico and {Yang}, Dongxin and {Citraro}, Saverio and {Nasimi}, Hikmat and {Yu}, Jiandong and {Jin}, Ge and {Zeng}, Ming and {An}, Peng and {Baldini}, Luca and {Bellazzini}, Ronaldo and {Brez}, Alessandro and {Latronico}, Luca and {Sgr{\`o}}, Carmelo and {Spandre}, Gloria and {Pinchera}, Michele and {Muleri}, Fabio and {Soffitta}, Paolo},
        title = "{X-Ray Polarimetry of the Crab Nebula with PolarLight: Polarization Recovery after the Glitch and a Secular Position Angle Variation}",
      journal = {\apjl},
         year = 2021,
        month = may,
       volume = {912},
       number = {2},
          eid = {L28},
        pages = {L28},
          doi = {10.3847/2041-8213/abfb00},
archivePrefix = {arXiv},
       eprint = {2104.11391},
 primaryClass = {astro-ph.HE},
       adsurl = {https://ui.adsabs.harvard.edu/abs/2021ApJ...912L..28L}
}

@ARTICLE{Long2023,
       author = {{Long}, Xiangyun and {Feng}, Hua and {Li}, Hong and {Kong}, Ling-Da and {Heyl}, Jeremy and {Ji}, Long and {Tao}, Lian and {Muleri}, Fabio and {Wu}, Qiong and {Zhu}, Jiahuan and {Huang}, Jiahui and {Minuti}, Massimo and {Jiang}, Weichun and {Citraro}, Saverio and {Nasimi}, Hikmat and {Yu}, Jiandong and {Jin}, Ge and {Zeng}, Ming and {An}, Peng and {Baldini}, Luca and {Bellazzini}, Ronaldo and {Brez}, Alessandro and {Latronico}, Luca and {Sgr{\`o}}, Carmelo and {Spandre}, Gloria and {Pinchera}, Michele and {Soffitta}, Paolo and {Costa}, Enrico},
        title = "{X-Ray Polarimetry of the Accreting Pulsar 1A 0535+262 in the Supercritical State with PolarLight}",
      journal = {\apj},
         year = 2023,
        month = jun,
       volume = {950},
       number = {2},
          eid = {76},
        pages = {76},
          doi = {10.3847/1538-4357/acd0af},
archivePrefix = {arXiv},
       eprint = {2304.13599},
 primaryClass = {astro-ph.HE},
       adsurl = {https://ui.adsabs.harvard.edu/abs/2023ApJ...950...76L}
}

@ARTICLE{Long2022,
       author = {{Long}, Xiangyun and {Feng}, Hua and {Li}, Hong and {Zhu}, Jiahuan and {Wu}, Qiong and {Huang}, Jiahui and {Minuti}, Massimo and {Jiang}, Weichun and {Yang}, Dongxin and {Citraro}, Saverio and {Nasimi}, Hikmat and {Yu}, Jiandong and {Jin}, Ge and {Zeng}, Ming and {An}, Peng and {Jiang}, Jiachen and {Costa}, Enrico and {Baldini}, Luca and {Bellazzini}, Ronaldo and {Brez}, Alessandro and {Latronico}, Luca and {Sgr{\`o}}, Carmelo and {Spandre}, Gloria and {Pinchera}, Michele and {Muleri}, Fabio and {Soffitta}, Paolo},
        title = "{A Significant Detection of X-ray Polarization in Sco X-1 with PolarLight and Constraints on the Corona Geometry}",
      journal = {\apjl},
         year = 2022,
        month = jan,
       volume = {924},
       number = {1},
          eid = {L13},
        pages = {L13},
          doi = {10.3847/2041-8213/ac4673},
archivePrefix = {arXiv},
       eprint = {2112.02837},
 primaryClass = {astro-ph.HE},
       adsurl = {https://ui.adsabs.harvard.edu/abs/2022ApJ...924L..13L}
}

@ARTICLE{Luo2021,
       author = {{Luo}, Jing and {Ransom}, Scott and {Demorest}, Paul and {Ray}, Paul S. and {Archibald}, Anne and {Kerr}, Matthew and {Jennings}, Ross J. and {Bachetti}, Matteo and {van Haasteren}, Rutger and {Champagne}, Chloe A. and {Colen}, Jonathan and {Phillips}, Camryn and {Zimmerman}, Josef and {Stovall}, Kevin and {Lam}, Michael T. and {Jenet}, Fredrick A.},
        title = "{PINT: A Modern Software Package for Pulsar Timing}",
      journal = {\apj},
         year = 2021,
        month = apr,
       volume = {911},
       number = {1},
          eid = {45},
        pages = {45},
          doi = {10.3847/1538-4357/abe62f},
archivePrefix = {arXiv},
       eprint = {2012.00074},
 primaryClass = {astro-ph.IM},
       adsurl = {https://ui.adsabs.harvard.edu/abs/2021ApJ...911...45L}
}

@ARTICLE{Manchester2005,
       author = {{Manchester}, R.~N. and {Hobbs}, G.~B. and {Teoh}, A. and {Hobbs}, M.},
        title = "{The Australia Telescope National Facility Pulsar Catalogue}",
      journal = {\aj},
         year = 2005,
        month = apr,
       volume = {129},
       number = {4},
        pages = {1993-2006},
          doi = {10.1086/428488},
archivePrefix = {arXiv},
       eprint = {astro-ph/0412641},
 primaryClass = {astro-ph},
       adsurl = {https://ui.adsabs.harvard.edu/abs/2005AJ....129.1993M}
}

@ARTICLE{Radhakrishnan1969b,
       author = {{Radhakrishnan}, V. and {Manchester}, R.~N.},
        title = "{Detection of a Change of State in the Pulsar PSR 0833-45}",
      journal = {\nat},
         year = 1969,
        month = apr,
       volume = {222},
       number = {5190},
        pages = {228-229},
          doi = {10.1038/222228a0},
       adsurl = {https://ui.adsabs.harvard.edu/abs/1969Natur.222..228R}
}

@ARTICLE{Ramsey2022,
       author = {{Ramsey}, Brian D. and {Bongiorno}, Stephen D. and {Kolodziejczak}, Jeffery J. and {Kilaru}, Kiranmayee and {Alexander}, Cheryl and {Baumgartner}, Wayne H. and {Breeding}, Shawn and {Elsner}, Ronald F. and {Le Roy}, Shelley and {McCracken}, Jeff and {Mitsuishi}, Ikuyuki and {O'Dell}, Stephen L. and {Pavelitz}, Steven D. and {Ranganathan}, Jaganathan and {Sanchez}, Javier and {Speegle}, Chet O. and {Thomas}, Nicholas and {Weddendorf}, Bruce and {Weisskopf}, Martin C.},
        title = "{Optics for the imaging x-ray polarimetry explorer}",
      journal = {Journal of Astronomical Telescopes, Instruments, and Systems},
         year = 2022,
        month = apr,
       volume = {8},
          eid = {024003},
        pages = {024003},
          doi = {10.1117/1.JATIS.8.2.024003},
       adsurl = {https://ui.adsabs.harvard.edu/abs/2022JATIS...8b4003R}
}

@ARTICLE{Ng2004,
       author = {{Ng}, C.-Y. and {Romani}, Roger W.},
        title = "{Fitting Pulsar Wind Tori}",
      journal = {\apj},
         year = 2004,
        month = jan,
       volume = {601},
       number = {1},
        pages = {479-484},
          doi = {10.1086/380486},
archivePrefix = {arXiv},
       eprint = {astro-ph/0310155},
 primaryClass = {astro-ph},
       adsurl = {https://ui.adsabs.harvard.edu/abs/2004ApJ...601..479N}
}

@MISC{Ruderman1998,
       author = {{Ruderman}, Malvin and {Zhu}, Tianhua and {Chen}, Kaiyou},
        title = "{Neutron Star Magnetic Field Evolution, Crust Movement, and Glitches: Erratum}",
         year = 1998,
        month = aug,
        pages = {1027},
          doi = {10.1086/305952},
    publisher = {IOP},
       adsurl = {https://ui.adsabs.harvard.edu/abs/1998ApJ...502.1027R}
}

@ARTICLE{Shaw2019,
       author = {{Shaw}, Benjamin and {Mickaliger}, Mitchell and {Keith}, Michael and {Lyne}, Andrew and {Stappers}, Benjamin and {Weltevrede}, Patrick},
        title = "{A glitch in the Crab pulsar (PSR B0531+21)}",
      journal = {The Astronomer's Telegram},
         year = 2019,
        month = jul,
       volume = {12957},
        pages = {1},
       adsurl = {https://ui.adsabs.harvard.edu/abs/2019ATel12957....1S}
}

@ARTICLE{Shaw2025,
       author = {{Shaw}, Benjamin and {Antonopoulou}, Danai and {Keith}, Michael J. and {Lyne}, Andrew G. and {Mickaliger}, Mitchell B. and {Stappers}, Benjamin W. and {Weltevrede}, Patrick},
        title = "{A glitch in the Crab pulsar (PSR B0531+21)}",
      journal = {The Astronomer's Telegram},
         year = 2025,
        month = jul,
       volume = {17298},
        pages = {1},
       adsurl = {https://ui.adsabs.harvard.edu/abs/2025ATel17298....1S}
}

@ARTICLE{Shaw2025b,
       author = {{Shaw}, Benjamin and {Antonopoulou}, Danai and {Keith}, Michael J. and {Lyne}, Andrew G. and {Mickaliger}, Mitchell B. and {Stappers}, Benjamin W. and {Weltevrede}, Patrick},
        title = "{Another glitch in the Crab pulsar (PSR B0531+21) following the July 2025 event}",
      journal = {The Astronomer's Telegram},
         year = 2025,
        month = aug,
       volume = {17331},
        pages = {1},
       adsurl = {https://ui.adsabs.harvard.edu/abs/2025ATel17331....1S}
}

@ARTICLE{Shaw2021,
       author = {{Shaw}, B. and {Keith}, M.~J. and {Lyne}, A.~G. and {Mickaliger}, M.~B. and {Stappers}, B.~W. and {Turner}, J.~D. and {Weltevrede}, P.},
        title = "{The slow rise and recovery of the 2019 Crab pulsar glitch}",
      journal = {\mnras},
         year = 2021,
        month = jul,
       volume = {505},
       number = {1},
        pages = {L6-L10},
          doi = {10.1093/mnrasl/slab038},
archivePrefix = {arXiv},
       eprint = {2103.13180},
 primaryClass = {astro-ph.HE},
       adsurl = {https://ui.adsabs.harvard.edu/abs/2021MNRAS.505L...6S}
}

@ARTICLE{Slowikowska2009,
       author = {{S{\l}owikowska}, A. and {Kanbach}, G. and {Kramer}, M. and {Stefanescu}, A.},
        title = "{Optical polarization of the Crab pulsar: precision measurements and comparison to the radio emission}",
      journal = {\mnras},
         year = 2009,
        month = jul,
       volume = {397},
       number = {1},
        pages = {103-123},
          doi = {10.1111/j.1365-2966.2009.14935.x},
archivePrefix = {arXiv},
       eprint = {0901.4559},
 primaryClass = {astro-ph.SR},
       adsurl = {https://ui.adsabs.harvard.edu/abs/2009MNRAS.397..103S}
}

@ARTICLE{Soffitta2013,
   author = {{Soffitta}, P. and {Muleri}, F. and {Fabiani}, S. and {Costa}, E. and
    {Bellazzini}, R. and {Brez}, A. and {Minuti}, M. and {Pinchera}, M. and
    {Spandre}, G.},
    title = "{Measurement of the position resolution of the Gas Pixel Detector}",
  journal = {Nuclear Instruments and Methods in Physics Research A},
archivePrefix = "arXiv",
   eprint = {1208.6330},
 primaryClass = "astro-ph.IM",
     year = 2013,
    month = feb,
   volume = 700,
    pages = {99-105},
      doi = {10.1016/j.nima.2012.09.055},
   adsurl = {http://adsabs.harvard.edu/abs/2013NIMPA.700...99S}
}

@ARTICLE{Soffitta2021,
       author = {{Soffitta}, Paolo and {Baldini}, Luca and {Bellazzini}, Ronaldo and {Costa}, Enrico and {Latronico}, Luca and {Muleri}, Fabio and {Del Monte}, Ettore and {Fabiani}, Sergio and {Minuti}, Massimo and {Pinchera}, Michele and {Sgro'}, Carmelo and {Spandre}, Gloria and {Trois}, Alessio and {Amici}, Fabrizio and {Andersson}, Hans and {Attina'}, Primo and {Bachetti}, Matteo and {Barbanera}, Mattia and {Borotto}, Fabio and {Brez}, Alessandro and {Brienza}, Daniele and {Caporale}, Ciro and {Cardelli}, Claudia and {Carpentiero}, Rita and {Castellano}, Simone and {Castronuovo}, Marco and {Cavalli}, Luca and {Cavazzuti}, Elisabetta and {Ceccanti}, Marco and {Centrone}, Mauro and {Ciprini}, Stefano and {Citraro}, Saverio and {D'Amico}, Fabio and {D'Alba}, Elisa and {Di Cosimo}, Sergio and {Di Lalla}, Niccolo' and {Di Marco}, Alessandro and {Di Persio}, Giuseppe and {Donnarumma}, Immacolata and {Evangelista}, Yuri and {Ferrazzoli}, Riccardo and {Hayato}, Asami and {Kitaguchi}, Takao and {La Monaca}, Fabio and {Lefevre}, Carlo and {Loffredo}, Pasqualino and {Lorenzi}, Paolo and {Lucchesi}, Leonardo and {Magazzu}, Carlo and {Maldera}, Simone and {Manfreda}, Alberto and {Mangraviti}, Elio and {Marengo}, Marco and {Matt}, Giorgio and {Mereu}, Paolo and {Morbidini}, Alfredo and {Mosti}, Federico and {Nakano}, Toshio and {Nasimi}, Hikmat and {Negri}, Barbara and {Nenonen}, Seppo and {Nuti}, Alessio and {Orsini}, Leonardo and {Perri}, Matteo and {Pesce-Rollins}, Melissa and {Piazzolla}, Raffaele and {Pilia}, Maura and {Profeti}, Alessandro and {Puccetti}, Simonetta and {Rankin}, John and {Ratheesh}, Ajay and {Rubini}, Alda and {Santoli}, Francesco and {Sarra}, Paolo and {Scalise}, Emanuele and {Sciortino}, Andrea and {Tamagawa}, Toru and {Tardiola}, Marcello and {Tobia}, Antonino and {Vimercati}, Marco and {Xie}, Fei},
        title = "{The Instrument of the Imaging X-Ray Polarimetry Explorer}",
      journal = {\aj},
         year = 2021,
        month = nov,
       volume = {162},
       number = {5},
          eid = {208},
        pages = {208},
          doi = {10.3847/1538-3881/ac19b0},
archivePrefix = {arXiv},
       eprint = {2108.00284},
 primaryClass = {astro-ph.IM},
       adsurl = {https://ui.adsabs.harvard.edu/abs/2021AJ....162..208S}
}

@ARTICLE{Weisskopf2011,
       author = {{Weisskopf}, Martin C. and {Tennant}, Allyn F. and {Yakovlev}, Dmitry G. and {Harding}, Alice and {Zavlin}, Vyacheslav E. and {O'Dell}, Stephen L. and {Elsner}, Ronald F. and {Becker}, Werner},
        title = "{Chandra Phase-resolved X-Ray Spectroscopy of the Crab Pulsar}",
      journal = {\apj},
         year = 2011,
        month = dec,
       volume = {743},
       number = {2},
          eid = {139},
        pages = {139},
          doi = {10.1088/0004-637X/743/2/139},
archivePrefix = {arXiv},
       eprint = {1106.3270},
 primaryClass = {astro-ph.GA},
       adsurl = {https://ui.adsabs.harvard.edu/abs/2011ApJ...743..139W}
}

@ARTICLE{Weisskopf2022,
       author = {{Weisskopf}, Martin C. and {Soffitta}, Paolo and {Baldini}, Luca and {Ramsey}, Brian D. and {O'Dell}, Stephen L. and {Romani}, Roger W. and {Matt}, Giorgio and {Deininger}, William D. and {Baumgartner}, Wayne H. and {Bellazzini}, Ronaldo and {Costa}, Enrico and {Kolodziejczak}, Jeffery J. and {Latronico}, Luca and {Marshall}, Herman L. and {Muleri}, Fabio and {Bongiorno}, Stephen D. and {Tennant}, Allyn and {Bucciantini}, Niccolo and {Dovciak}, Michal and {Marin}, Frederic and {Marscher}, Alan and {Poutanen}, Juri and {Slane}, Pat and {Turolla}, Roberto and {Kalinowski}, William and {Di Marco}, Alessandro and {Fabiani}, Sergio and {Minuti}, Massimo and {La Monaca}, Fabio and {Pinchera}, Michele and {Rankin}, John and {Sgro'}, Carmelo and {Trois}, Alessio and {Xie}, Fei and {Alexander}, Cheryl and {Allen}, D. Zachery and {Amici}, Fabrizio and {Andersen}, Jason and {Antonelli}, Angelo and {Antoniak}, Spencer and {Attin{\`a}}, Primo and {Barbanera}, Mattia and {Bachetti}, Matteo and {Baggett}, Randy M. and {Bladt}, Jeff and {Brez}, Alessandro and {Bonino}, Raffaella and {Boree}, Christopher and {Borotto}, Fabio and {Breeding}, Shawn and {Brienza}, Daniele and {Bygott}, H. Kyle and {Caporale}, Ciro and {Cardelli}, Claudia and {Carpentiero}, Rita and {Castellano}, Simone and {Castronuovo}, Marco and {Cavalli}, Luca and {Cavazzuti}, Elisabetta and {Ceccanti}, Marco and {Centrone}, Mauro and {Citraro}, Saverio and {D'Amico}, Fabio and {D'Alba}, Elisa and {Di Gesu}, Laura and {Del Monte}, Ettore and {Dietz}, Kurtis L. and {Di Lalla}, Niccolo' and {Persio}, Giuseppe Di and {Dolan}, David and {Donnarumma}, Immacolata and {Evangelista}, Yuri and {Ferrant}, Kevin and {Ferrazzoli}, Riccardo and {Ferrie}, MacKenzie and {Footdale}, Joseph and {Forsyth}, Brent and {Foster}, Michelle and {Garelick}, Benjamin and {Gunji}, Shuichi and {Gurnee}, Eli and {Head}, Michael and {Hibbard}, Grant and {Johnson}, Samantha and {Kelly}, Erik and {Kilaru}, Kiranmayee and {Lefevre}, Carlo and {Roy}, Shelley Le and {Loffredo}, Pasqualino and {Lorenzi}, Paolo and {Lucchesi}, Leonardo and {Maddox}, Tyler and {Magazzu}, Guido and {Maldera}, Simone and {Manfreda}, Alberto and {Mangraviti}, Elio and {Marengo}, Marco and {Marrocchesi}, Alessandra and {Massaro}, Francesco and {Mauger}, David and {McCracken}, Jeffrey and {McEachen}, Michael and {Mize}, Rondal and {Mereu}, Paolo and {Mitchell}, Scott and {Mitsuishi}, Ikuyuki and {Morbidini}, Alfredo and {Mosti}, Federico and {Nasimi}, Hikmat and {Negri}, Barbara and {Negro}, Michela and {Nguyen}, Toan and {Nitschke}, Isaac and {Nuti}, Alessio and {Onizuka}, Mitch and {Oppedisano}, Chiara and {Orsini}, Leonardo and {Osborne}, Darren and {Pacheco}, Richard and {Paggi}, Alessandro and {Painter}, Will and {Pavelitz}, Steven D. and {Pentz}, Christina and {Piazzolla}, Raffaele and {Perri}, Matteo and {Pesce-Rollins}, Melissa and {Peterson}, Colin and {Pilia}, Maura and {Profeti}, Alessandro and {Puccetti}, Simonetta and {Ranganathan}, Jaganathan and {Ratheesh}, Ajay and {Reedy}, Lee and {Root}, Noah and {Rubini}, Alda and {Ruswick}, Stephanie and {Sanchez}, Javier and {Sarra}, Paolo and {Santoli}, Francesco and {Scalise}, Emanuele and {Sciortino}, Andrea and {Schroeder}, Christopher and {Seek}, Tim and {Sosdian}, Kalie and {Spandre}, Gloria and {Speegle}, Chet O. and {Tamagawa}, Toru and {Tardiola}, Marcello and {Tobia}, Antonino and {Thomas}, Nicholas E. and {Valerie}, Robert and {Vimercati}, Marco and {Walden}, Amy L. and {Weddendorf}, Bruce and {Wedmore}, Jeffrey and {Welch}, David and {Zanetti}, Davide and {Zanetti}, Francesco},
        title = "{The Imaging X-Ray Polarimetry Explorer (IXPE): Pre-Launch}",
      journal = {Journal of Astronomical Telescopes, Instruments, and Systems},
         year = 2022,
        month = apr,
       volume = {8},
       number = {2},
          eid = {026002},
        pages = {026002},
          doi = {10.1117/1.JATIS.8.2.026002},
archivePrefix = {arXiv},
       eprint = {2112.01269},
 primaryClass = {astro-ph.IM},
       adsurl = {https://ui.adsabs.harvard.edu/abs/2022JATIS...8b6002W}
}

@ARTICLE{Wong2023,
       author = {{Wong}, Josephine and {Romani}, Roger W. and {Dinsmore}, Jack T.},
        title = "{Improved Measurements of the IXPE Crab Polarization}",
      journal = {\apj},
         year = 2023,
        month = aug,
       volume = {953},
       number = {1},
          eid = {28},
        pages = {28},
          doi = {10.3847/1538-4357/acdc1d},
archivePrefix = {arXiv},
       eprint = {2306.08788},
 primaryClass = {astro-ph.HE},
       adsurl = {https://ui.adsabs.harvard.edu/abs/2023ApJ...953...28W}
}

@ARTICLE{Gonzalez2025,
       author = {{Gonz{\'a}lez-Caniulef}, Denis and {Heyl}, Jeremy and {Fabiani}, Sergio and {Soffitta}, Paolo and {Costa}, Enrico and {Bucciantini}, Niccol{\`o} and {Kirmizibayrak}, Demet and {Xie}, Fei},
        title = "{Crab pulsar: IXPE observations reveal unified polarization properties in the optical and soft X-ray bands}",
      journal = {\aap},
         year = 2025,
        month = jan,
       volume = {693},
          eid = {A152},
        pages = {A152},
          doi = {10.1051/0004-6361/202451815},
archivePrefix = {arXiv},
       eprint = {2408.03245},
 primaryClass = {astro-ph.HE},
       adsurl = {https://ui.adsabs.harvard.edu/abs/2025A&A...693A.152G}
}

@ARTICLE{Wong2024,
       author = {{Wong}, Josephine and {Mizuno}, Tsunefumi and {Bucciantini}, Niccol{\'o} and {Romani}, Roger W. and {Yang}, Yi-Jung and {Liu}, Kuan and {Deng}, Wei and {Goya}, Kazuho and {Xie}, Fei and {Pilia}, Maura and {Kaaret}, Philip and {Weisskopf}, Martin C. and {Silvestri}, Stefano and {Ng}, C. -Y. and {Chen}, Chien-Ting and {Agudo}, Iv{\'a}n and {Antonelli}, Lucio A. and {Bachetti}, Matteo and {Baldini}, Luca and {Baumgartner}, Wayne H. and {Bellazzini}, Ronaldo and {Bianchi}, Stefano and {Bongiorno}, Stephen D. and {Bonino}, Raffaella and {Brez}, Alessandro and {Capitanio}, Fiamma and {Castellano}, Simone and {Cavazzuti}, Elisabetta and {Ciprini}, Stefano and {Costa}, Enrico and {De Rosa}, Alessandra and {Del Monte}, Ettore and {Di Gesu}, Laura and {Di Lalla}, Niccol{\'o} and {Di Marco}, Alessandro and {Donnarumma}, Immacolata and {Doroshenko}, Victor and {Dov{\v{c}}iak}, Michal and {Ehlert}, Steven R. and {Enoto}, Teruaki and {Evangelista}, Yuri and {Fabiani}, Sergio and {Ferrazzoli}, Riccardo and {Garcia}, Javier A. and {Gunji}, Shuichi and {Heyl}, Jeremy and {Iwakiri}, Wataru and {Jorstad}, Svetlana G. and {Karas}, Vladimir and {Kislat}, Fabian and {Kitaguchi}, Takao and {Kolodziejczak}, Jeffery J. and {Krawczynski}, Henric and {La Monaca}, Fabio and {Latronico}, Luca and {Liodakis}, Ioannis and {Maldera}, Simone and {Manfreda}, Alberto and {Marin}, Fr{\'e}d{\'e}ric and {Marinucci}, Andrea and {Marscher}, Alan P. and {Marshall}, Herman L. and {Massaro}, Francesco and {Matt}, Giorgio and {Mitsuishi}, Ikuyuki and {Muleri}, Fabio and {Negro}, Michela and {O'Dell}, Stephen L. and {Omodei}, Nicola and {Oppedisano}, Chiara and {Papitto}, Alessandro and {Pavlov}, George G. and {Peirson}, Abel Lawrence and {Perri}, Matteo and {Pesce-Rollins}, Melissa and {Petrucci}, Pierre-Olivier and {Possenti}, Andrea and {Poutanen}, Juri and {Puccetti}, Simonetta and {Ramsey}, Brian D. and {Rankin}, John and {Ratheesh}, Ajay and {Roberts}, Oliver J. and {Sgr{\'o}}, Carmelo and {Slane}, Patrick and {Soffitta}, Paolo and {Spandre}, Gloria and {Swartz}, Douglas A. and {Tamagawa}, Toru and {Tavecchio}, Fabrizio and {Taverna}, Roberto and {Tawara}, Yuzuru and {Tennant}, Allyn F. and {Thomas}, Nicholas E. and {Tombesi}, Francesco and {Trois}, Alessio and {Tsygankov}, Sergey and {Turolla}, Roberto and {Vink}, Jacco and {Wu}, Kinwah and {Zane}, Silvia},
        title = "{Analysis of Crab X-Ray Polarization Using Deeper Imaging X-Ray Polarimetry Explorer Observations}",
      journal = {\apj},
         year = 2024,
        month = oct,
       volume = {973},
       number = {2},
          eid = {172},
        pages = {172},
          doi = {10.3847/1538-4357/ad6309},
archivePrefix = {arXiv},
       eprint = {2407.12779},
 primaryClass = {astro-ph.HE},
       adsurl = {https://ui.adsabs.harvard.edu/abs/2024ApJ...973..172W}
}

@ARTICLE{Ferrazzoli2025,
       author = {{Ferrazzoli}, Riccardo and {Costa}, Enrico and {Fabiani}, Sergio and {Kaaret}, Philip and {O'Dell}, Stephen L. and {Ramsey}, Brian D. and {Soffitta}, Paolo and {Baldini}, Luca and {Bellazzini}, Ronaldo and {Di Marco}, Alessandro and {La Monaca}, Fabio and {Latronico}, Luca and {Manfreda}, Alberto and {Muleri}, Fabio and {Rankin}, John and {Sgr{\'o}}, Carmelo and {Silvestri}, Stefano and {Weisskopf}, Martin C.},
        title = "{In-flight Performance of the IXPE Telescopes}",
      journal = {\aj},
         year = 2025,
        month = dec,
       volume = {170},
       number = {6},
          eid = {325},
        pages = {325},
          doi = {10.3847/1538-3881/ae11a5},
archivePrefix = {arXiv},
       eprint = {2510.06963},
 primaryClass = {astro-ph.IM},
       adsurl = {https://ui.adsabs.harvard.edu/abs/2025AJ....170..325F}
}
\bibliographystyle{aasjournalv7}



\end{document}